\documentclass[twocolumn]{aastex631}

\usepackage{amsmath}
\usepackage{tabularx}
\usepackage{rotating}

\newcommand{\DCOS}{DCOs\,}

\newcommand{\bham}{\affiliation{Institute for Gravitational Wave Astronomy \& School of Physics and Astronomy, University of Birmingham, Birmingham, United Kingdom}}

\newcommand{\palma}{\affiliation{Departament de F\'isica, Universitat de les Illes Balears, IAC3--IEEC, Crta. Valldemossa km 7.5, E-07122 Palma, Spain}}

\newcommand{\cca}{\affiliation{Center for Computational Astrophysics, Flatiron Institute, 162 5th Avenue, New York, NY 10010}}

\begin{document}

\title{
Identifying galactic binary systems of neutron stars and black holes with LISA
}

\author[0000-0001-5532-3622]{Hannah~Middleton}\thanks{hannahm@star.sr.bham.ac.uk}
\bham

\author[0000-0002-6719-8686]{Panagiota~Kolitsidou}\thanks{panagiota.kolitsidou@uib.es}
\palma
\bham

\author[0000-0001-5438-9152]{Antoine~Klein}
\bham

\author[0009-0006-9980-3437]{Connor~Adam}
\bham

\author[0009-0009-7374-3722]{Rowan~Chalmers}
\bham

\author[0000-0002-6254-1617]{Alberto~Vecchio}\thanks{a.vecchio@bham.ac.uk}
\bham
\cca

\begin{abstract}
The Laser Interferometer Space Antenna (LISA) will detect $\sim 100$ Galactic binary systems comprised of black holes and neutron stars. Distinguishing these binaries from the $\sim 10^4$ double white dwarfs detectable by LISA will be challenging. In the absence of any other information, the inferred component masses can be used to classify the nature of these objects. However, short-period Galactic binaries $\sim 10^7$--$10^3\,\mathrm{yr}$ from coalescence produce a quasi-monochromatic signal which carries little information about their masses.  We generate synthetic LISA datasets containing gravitational waves from Galactic binary black holes, binary neutron stars, and black hole--neutron stars drawn from an astrophysically realistic population produced through the isolated binary evolution channel. We use a Bayesian inference pipeline to explore the accuracy with which the component masses can be measured. LISA will be able to infer component masses for $\approx 10\%$--$50\%$ of the detected systems by measuring the orbital eccentricity, periapse precession frequency, and gravitational-wave-induced frequency derivative. 
Typical fractional mass errors are $\approx 1\%$--$100\%$ (depending on the specific value of the source parameters) enabling the classification of the constituent objects as black holes or neutron stars. For these binaries, LISA will also be able to determine their 3-dimensional positions in the Milky Way. LISA detections of these double compact objects (DCOs) will provide new insights to their astrophysical formation processes, the Galactic population, and the Milky Way's star formation history. 
However, for a significant fraction of the LISA-detected binaries the nature of their constituent objects may remain unclear.

\end{abstract}


\section{Introduction} 
\label{sec:intro}

The Galactic population of neutron stars (NSs) and black holes (BHs) is currently known through radio~\citep{2005AJ....129.1993M}, X-ray~\citep{2006ARA&A..44...49R, 2014SSRv..183..223C, 2021NewAR..9301618M}, astrometric~\citep{2023MNRAS.518.1057E, 2023MNRAS.521.4323E, 2024A&A...686L...2G, 2024OJAp....7E..27E, 2024OJAp....7E..58E} and micro-lensing~\citep{2016ApJ...830...41L, 2020A&A...636A..20W} observations. 
These surveys are building a picture of the populations of these objects in the Milky Way, their environments, formation, and evolution pathways. 
However, they also suffer from selection effects that limit the ability to fully probe the complex physical processes at play.

The Laser Interferometer Space Antenna~\citep[LISA;][]{2017arXiv170200786A, 2023LRR....26....2A, 2024arXiv240207571C} will offer a new means to probe the Galactic (and Local Group) population of binary black holes (BHBHs), binary neutron stars (NSNSs) and neutron star-black holes (BHNSs) through gravitational waves (GWs). 
LISA observations do not suffer from limitations of identifying binaries in dense stellar fields and/or through the Galactic plane. 
In addition, LISA selection effects are different from those affecting electromagnetic surveys, and are rather simple to quantify.

Hereafter, we will refer to BHBHs, BHNSs, NSNSs as double compact objects (\DCOS); with slight abuse of language, our nomenclature excludes binary systems in which one or both constituents are a white dwarf (WD) from \DCOS.

Many studies, considering either isolated binary evolution or dynamical formation channels suggest that LISA will detect between a few and hundreds of \DCOS~\citep{2001A&A...375..890N, 
2018MNRAS.480.2704L, 
2018MNRAS.481.4775D, 
2018PhRvL.120s1103K,
2019PhRvD.100d3010S, 
2020MNRAS.494L..75S, 
2020ApJ...892L...9A, 
2020ApJ...898...71B,
2020MNRAS.492.3061L,
2022ApJ...937..118W, 
2024MNRAS.534.1707T,
2025arXiv250118682X}, 
during its mission lifetime, nominally consisting of a science-mode observing period of $4.5\,\mathrm{yr}$ with the opportunity to extend it to $10\,\mathrm{yr}$. 
Such a GW-selected sample of \DCOS will provide an important complement to objects discovered via other surveys. 
It also provides a sample of systems in the local Universe that can be compared to the population of \textit{coalescing} binaries detected by the ground-based network of GW laser interferometers~\citep{GWTC1, GWTC2, GWTC3, GW230529}, which, on the timescale of LISA science operations, will reach cosmological distances \citep[\textit{e.g.}][and references therein]{2021ApJ...913L...5N, ET-BLUEBOOK}. 
These \DCOS will directly inform the formation pathways of these binaries, binary evolution processes, the star formation rate and chemical composition of our Galaxy and the distribution of objects in the Milky Way.

BHBHs, BHNSs and NSNSs will represent $\sim 0.1\%$--$1\%$ of the total number of LISA detected stellar-mass compact binaries with periods $\sim 1\,\mathrm{min}$--$1\,\mathrm{hr}$. Binaries containing WDs will be much more abundant; in fact LISA is forecast to observe $\sim 10^4$ double WDs~\citep{2001A&A...375..890N, 
2020ApJ...898...71B, 2020ApJ...894L..15R, 2020A&A...638A.153K, 2023LRR....26....2A, 2025ApJ...981...66D} and $\sim 10^3$ neutron star-white dwarfs~\citep{2020ApJ...898...71B, 2024MNRAS.530..844K} in the Galaxy and the Local Group.
As such WD systems have been studied elsewhere, we restrict our study to only BHBHs, BHNSs, and NSNS. 
However the issue of distinguishing between systems containing NS and WD is an important one which will be further discussed in Section~\ref{sec:analysis_measureability}.

The key question we want to explore in this paper is to what extent one can measure the mass of the \textit{individual} components of these \DCOS to enable the identification of this small but precious sample of Galactic NSs and BHs, and distinguish these binaries from the overwhelming majority of Galactic systems consisting of double WDs.

All Galactic short-period binaries observed by LISA are expected to be on sufficiently wide orbits that their orbital frequency hardly changes due to radiation reaction during the LISA observing period.
Regardless of the nature of the objects, these binaries therefore produce in the LISA time series a quasi-monochromatic signal slowly drifting up in frequency with additional modulations due to the motion of the LISA constellation~\citep{1998PhRvD..57.7089C}. 
The key signatures on the phase evolution produced by masses (and spins, and, where relevant, tidal effects) upon which high-frequency observations rely to measure these quantities are mostly lost~\citep[see \textit{e.g.}][and references therein]{2024LRR....27....4B}.

Assuming there are no additional physical processes affecting the orbital evolution of a binary other than gravitational radiation reaction (\textit{e.g.}, mass transfer and/or tidal effects that are of importance for binaries containing WDs) it is well known that two independent mass parameters can still be measured if one can measure the frequency derivative of the signal---which provides a direct measure of the chirp mass of the system---the periastron advance---which provides a direct measurement of the total mass---and the eccentricity of the orbit---which affects the rate at which the frequency evolves and the periastron advances~\citep[see, \textit{e.g.},][]{2001PhRvL..87y1101S, 2024MNRAS.531.2817M}. 
These are the observational signatures that allow for the measurement of two independent mass parameters, and therefore the determination of the masses of the individual constituents of a binary.

In this paper, we consider an astrophysical population of Galactic binaries following~\cite{2022ApJ...937..118W} that reflects the current astrophysical and observational understanding of double compact objects. 
Through an end-to-end analysis of synthetic LISA data we explore LISA's ability to measure binary component masses of \DCOS, and therefore identify the constituent objects as NSs or BHs, and distinguish them from double WDs. 
We show that LISA will be able to measure the individual component masses of $\approx 50\%$ of the detected NSNSs and $\approx 10\%$ of the detected BHBHs and BHNSs with a fractional error $\sim 1\%$--$100\%$, depending on the specific mass, orbital period, eccentricity, and distance to the source. 
In these circumstances LISA will also be able to measure the orbital eccentricity down to $\approx 10^{-3}$ and determine the location of the binary in the Milky Way. 
The latter will enable follow-up observations of the environment hosting the binary, which could provide additional important clues about their formation pathway.

The paper is structured as follows: In Section~\ref{sec:signal} we summarise the properties of GWs emitted by \DCOS and the assumptions on the signal used in this paper; in Section~\ref{sec:LISAanalysis} we review the Bayesian analysis method used for this study; Section~\ref{sec:astro_model} contains a summary of the synthetic Galactic binary population model from which we draw binary systems to generate LISA data; in Section~\ref{sec:results} we present and discuss the results; and Section~\ref{sec:concl} contains our conclusions and a discussion of the implications on the analysis of the LISA data. 
The accompanying data release includes posterior chains and example notebooks to reproduce selected plots at \url{doi.org/10.5281/zenodo.15704561}.

\section{Observed signal} 
\label{sec:signal}

Here we briefly summarise the key concepts and equations that describe the signal produced by \DCOS at $\sim \mathrm{milliHertz}$ frequencies accessible to LISA.
We use the convention $G=c=1$ throughout.

\subsection{Gravitational waveforms}
\label{subsec:gw}

We consider a binary of component masses $m_{1,2}$ (here we use the convention $m_1 \ge m_2$), orbital period $P$, semi-major axis $a$, and orbital eccentricity $e$. 
We define the following mass parameters:
\begin{subequations}
\begin{align}
    M & = m_1 + m_2\,,
    \label{eq:M} \\
    \mu & = \frac{m_1 m_2}{M}\,,
    \label{eq:mu} \\
    {\cal M} & = M^{2/5}\mu^{3/5} = \eta^{3/5}\,M\,,
    \label{eq:mchirp} \\
    \eta & = \frac{\mu}{M}\,,
    \label{eq:eta} 
\end{align}
\end{subequations}
corresponding to the total mass, reduced mass, chirp mass, and symmetric mass ratio, respectively.

The leading post-Newtonian order time to coalescence is
\begin{equation}
    \tau = 0.2\,\left(\frac{\eta}{0.25}\right)^{-1}\,\left(\frac{P}{10^3\,\mathrm{s}}\right)^{8/3}\,\left(\frac{{\cal M}}{3\,M_\odot}\right)^{-5/3}\,u(e)\,\mathrm{Myr}\,,
    \label{eq:tau}
\end{equation}
where $u(e)$ is a function of eccentricity~\citep{1964PhRv..136.1224P}. 
For $(1 - e^2)\ll 1$, this function scales as $u \propto (1 - e^2)^{7/2}$; in the limit $e \rightarrow 0$, $u \rightarrow 1$. 
LISA's nominal science observing period is $4.5\,\mathrm{yr}$. 
Stellar-mass binaries in the $\mathrm{milliHertz}$ band are therefore far from coalescence and their orbital elements change slowly during the mission lifetime. 
As a consequence, one can model the gravitational radiation at the lowest post-Newtonian order and our notation closely follows~\cite{2024MNRAS.531.2817M}, which in turn is based on~\cite{1995MNRAS.274..115M}.

For a binary in circular orbit, at the leading post-Newtonian order, radiation is emitted at a frequency that is twice the orbital frequency. 
For a binary in an elliptical orbit---\DCOS in the LISA band will still have some residual eccentricity which is a signature of their evolutionary history~\citep[\textit{e.g.}][]{2020ApJ...898...71B, 2022ApJ...937..118W}---radiation is emitted at a (formally, infinite) number of harmonics. 
The two independent GW polarisation amplitudes can be written as
\begin{subequations}
\begin{align}
    h_+(t)  = &  -(1 + \cos^2{\iota}) \nonumber \\ 
    & \times \sum_{n=1}^N \mathcal{A}_n(e) \cos[2\pi f_n(t)\, t + \phi + (n-2)\Phi(t)]\,, 
    \label{eq:h+}\\
    h_\times(t) = & 2\cos{\iota} \nonumber \\
    & \times \sum_{n=1}^N \mathcal{A}_n(e) \sin[2\pi f_n(t)\, t + \phi + (n-2)\Phi(t)]\,,
    \label{eq:hx}
\end{align}
\end{subequations}
where $\iota$ is the orbit's inclination angle, which can be assumed constant throughout the year-long observation period, and $\phi$ is an arbitrary (constant) reference phase, that sets the phase of $h_+$ and $h_\times$ at some (arbitrary) reference time, \textit{e.g.}, the start of the observations. 
The frequencies $f_n (t)$ are defined in Eq.~(\ref{eq:f_n}) below.

The amplitudes of each harmonic entering Eqs.~(\ref{eq:h+}) and~(\ref{eq:hx}) are
\begin{equation}
    \mathcal{A}_n(e) = \mathcal{A} \,\left[\frac{S_n(e) + C_n(e)}{2(1-e^2)}\right]\,,
    \label{eq:An_e}
\end{equation}
where $S_n(e)$ and $C_n(e)$ are given by Eqs.~(A15) and (A16) in \cite{1995MNRAS.274..115M}, respectively. 
Eqs.~(\ref{eq:h+}) and~(\ref{eq:hx}) are an approximation that neglects the sub-dominant components $\propto [S_n(e) - C_n(e)]$. 
The overall (eccentricity independent) amplitude term in Eq.~(\ref{eq:An_e}) is
\begin{equation}
    \mathcal{A} = 2\,\frac{\mathcal{M}^{5/3}}{D} (\pi f_\mathrm{GW})^{2/3}\,.
    \label{eq:A}
\end{equation}
In the previous expression, $D$ is the (luminosity) distance to the source and $f_\mathrm{GW}$ is the gravitational frequency that is defined below in Eq.~(\ref{eq:fgw}). 
It is useful to note that in the limit $e\ll 1$, Eq.~(\ref{eq:An_e}) reduces to $\mathcal{A}_n(e) \approx 2 (\mathcal{A}/n)\, [g(n,e)]^{1/2}$, where $g(n,e)$ is given by Eq.~(20) in \cite{1963PhRv..131..435P} and is proportional to the power in the $n^\mathrm{th}$ harmonic. 
For a binary in circular orbit, the only non-zero amplitude term is $n=2$ and one recovers the usual result $\mathcal{A}_2(e = 0) = \mathcal{A}$.

The phase of the two waveform polarisations, Eqs.~(\ref{eq:h+}) and (\ref{eq:hx}), contain several terms that we now define. The frequencies $f_n$ are defined as
\begin{equation}
    f_n(t) = \frac{n}{2} (f_\mathrm{GW} + \frac{1}{2} \dot{f}_\mathrm{GW} t)\,,
    \label{eq:f_n}
\end{equation}
where 
\begin{equation}
    f_{\rm GW} = \frac{1}{\pi} \sqrt{\frac{M}{a^3}}\,,
    \label{eq:fgw}
\end{equation}
and the frequency derivative is
\begin{align}
    \dot{f}_\mathrm{GW} & = \frac{96}{5} \pi^{8/3} \mathcal{M}^{5/3} f_\mathrm{GW}^{11/3}\,F(e)\,,\\
    &= 5.8\times 10^{-18}\left(\frac{\mathcal{M}}{M_\odot}\right)^{5/3}\left(\frac{f_{\rm GW}}{1\,\mathrm{mHz}}\right)^{11/3}\,F(e)\,\mathrm{Hz}\,\mathrm{s}^{-1}.
    \label{eq:fgw_dot}
\end{align}
In Eq.~(\ref{eq:f_n}) we have taken into account the fact that the binary's orbital period changes on time-scales much longer than the duration of the observations, see Eq.~(\ref{eq:tau}), and have therefore modeled the frequency and phase evolution as a Taylor expansion, in which we retain the leading order term.

The function $F(e)$ in Eq.~(\ref{eq:fgw_dot}) is~\citep[][]{1963PhRv..131..435P}  
\begin{align} 
    F(e) = \frac{1+\frac{73}{24}e^2 + \frac{37}{96}e^4}{(1-e^2)^{7/2}}\,,
    \label{eq:F_e}
\end{align} 
and reflects the fact that a binary with non-zero eccentricity loses energy more efficiently through gravitational radiation than a binary in a circular orbit with the same period.

If a binary is in an eccentric orbit, its periastron advances. 
We define the periastron precession frequency as 
\begin{align}
    f_\mathrm{PP} & = 
    \frac{3\,\pi^{2/3}}{2 (1-e^2)}\,M^{2/3}\, f_\mathrm{GW}^{5/3}\\
    &= 9.3\times 10^{-9}\left(\frac{1}{1-e^2}\right)\left(\frac{M}{M_\odot}\right)^{2/3}\left(\frac{f_{\rm GW}}{1\,\mathrm{mHz}}\right)^{5/3}\,\mathrm{Hz}.
    \label{eq:fpp}
\end{align}
As a consequence, the argument of the periastron, $\Phi(t)$, does not remain constant throughout the observation, but advances according to
\begin{equation}
    \Phi(t) = \Phi_0 +2 \pi f_\mathrm{PP}t\,,
    \label{eq:Phi_t}
\end{equation}
where $\Phi_0$ is the reference value at some (arbitrary) reference time $t=0$.

\begin{table*}[htb!]
\centering
 \begin{tabular}{c c l l} 
 \hline
 parameter & description & signal generation & analysis prior range \\
 \hline
 $\mathcal{A}$         & amplitude                        & from W22 distribution                 & $\sim{\cal U}[\mathcal{A}_{\rm inj}/100, ~ 5\mathcal{A}_{\rm inj}]$    \\
 $\iota$               & inclination angle                & $\cos\iota \sim {\cal U}[-1, 1]$      & $\cos\iota \sim {\cal U}[-1, 1]$\\
 $\psi$                & polarisation angle               & $\sim {\cal U}[0, \pi]$               & $\sim{\cal U}[0,\pi]$   \\
 $l$                   & ecliptic longitude               & from W22 distribution                 & $\sim{\cal U}[0, 2\pi]$ \\
 $b$                   & ecliptic latitude                & from W22 distribution                 & $\sin b \sim {\cal U}[-1, 1]$ \\
 $\phi$                & GW initial phase                 & $\sim {\cal U}[0, 2\pi]$              & $\sim{\cal U}[0, 2\pi]$ \\
 $e$                   & orbital eccentricity             & from W22 distribution                 & \textbf{if} $e_{\rm inj} < 0.01$: $e \sim {\cal U}[0, 10 e_{\rm inj}]$\\
                       &                                  &                                       & \textbf{else:} $e\sim {\cal U}[0, \min(3 e_{\rm inj}, 0.9)]$ \\
 $\Phi_0$              & initial periastron argument      & $\sim {\cal U}[0, 2\pi]$              & $\sim {\cal U}[0, 2\pi]$ \\
 $f_\mathrm{GW}$       & GW frequency, Eq.~(\ref{eq:fgw}) & from W22 distribution                 &  $\sim {\cal U}[f_\mathrm{GW}-2/T_{\rm obs}, f_\mathrm{GW}+2/T_{\rm obs}]$ \\
 $\dot{f}_\mathrm{GW}$ & GW frequency derivative, Eq.~(\ref{eq:fgw_dot}) & from W22 distribution  &  $\sim {\cal U}[0, 5 \dot{f}_\mathrm{GW,inj}]$   \\
 ${f}_\mathrm{PP}$     & periapse, Eq (\ref{eq:fpp})      & from W22 distribution                 &  $\sim {\cal U}[0, 5 {f}_\mathrm{PP,inj}]$      \\
 \hline
 \end{tabular}
 \caption{The parameters that define the GW signal from a binary consisting of a black hole and/or a neutron star (BHBH, BHNS, NSNS) considered in this analysis. The signal generation column describes how the parameters are chosen to generate the signal to analyse. The ``W22 distribution" is the source parameters probability distribution released by~\cite{W22-data} for the sources detectable over a mission lifetime of $4\,\mathrm{yrs}$ in the fiducial model of~\cite{2022ApJ...937..118W}. The ``analysis prior range'' provides the functional form and range of the priors used in the analyses. The subscript ``$\mathrm{inj}$'' for some of the parameters refers to the value of the actual signal used in the data generation (the ``injection'' signal). In our analysis we use $T_\mathrm{obs} = 4\,\mathrm{yr}$.}
 \label{tab:param}
\end{table*}

\subsection{LISA observables}

The observables of the LISA observatory are the Time-Delay-Interferometry ~\citep[TDI; see][and references therein]{2021LRR....24....1T} time series that consist of combinations of the one-way phase measurements between the pairs of test masses hosted in the three spacecrafts of the LISA constellation, in which the spacecraft separation is $\approx 2.5\times 10^{6}\,\mathrm{km}$. 
The phase measurements carry the imprint of the gravitational waveforms, Eqs.~(\ref{eq:h+}) and~(\ref{eq:hx}).

For our analysis, we consider the usual three TDI output channels, conventionally named $A$, $E$, and $T$. 
We will use the short-hand notation $y_\alpha$ to label the three channels, with $\alpha = A, E, T$. 
We therefore formally define the three TDI time series as
\begin{equation}
    y_\alpha({\mathbf\theta}) = y_\alpha(h_+,h_\times, \psi, l, b)\,,
    \label{eq:y}
\end{equation}
where $\psi$ is the polarisation angle, and $(l, b)$ identifies the source location in the sky, through its ecliptic longitude and latitude, respectively. 
The LISA TDI observables are uniquely defined by $\psi$, $l$, and $b$ and any choice of eight independent parameters that fully determine $h_+(t)$ and $h_\times(t)$, according to the model discussed in Section~\ref{subsec:gw}. 
Here we use ${\mathbf \theta}$ to indicate the set of parameters that describe the observable signal. 
Our choice of signal parameters is summarised in Table~\ref{tab:param} and given by
\begin{equation}
    \mathbf{\theta} = \{\mathcal{A}, \iota, \psi, l, b, \phi, e, \Phi_0, f_\mathrm{GW}, \dot{f}_\mathrm{GW}, f_\mathrm{PP}\}\,.
    \label{eq:theta}
\end{equation}

\section{LISA data analysis}
\label{sec:LISAanalysis}

The goal of our analysis is to explore LISA's ability to measure the component masses of a binary.
We therefore want to compute the posterior probability density functions (PDFs) on the mass parameters from the PDFs of the observable parameters.

Given a dataset $y = \{y_\alpha; \alpha = A, E, T\}$, we need to compute the joint posterior PDF on the model parameters $\theta$, given by Eq.~(\ref{eq:theta}):
\begin{equation}
    p(\theta | y ) \propto {\cal L}(y|\theta) p(\theta)\,, 
    \label{eq:posterior}
\end{equation}
where ${\cal L}(y|\theta)$ is the likelihood function and $p(\theta)$ is the prior. 
The likelihood function is given by
\begin{equation}
    \mathcal{L}(d | \theta) \propto \prod_\alpha \exp\left(-\frac{1}{2}\langle d_\alpha - y_\alpha | d_\alpha - y_\alpha \rangle \right)\,,
    \label{eq:likelihood}
\end{equation}
where $d_\alpha = n_\alpha + y_\alpha$ is the data stream in each TDI channel. 
It contains contributions from noise $n_\alpha$, characterised by one-sided noise power spectral density (PSD) $S_\alpha(f)$ and the TDI outputs $y_\alpha$, which carry information about gravitational radiation. 
The likelihood is formulated under the usual assumption that $A$, $E$ and $T$ are noise orthogonal. 
During the mission operation this will not be exactly true~\citep{2021LRR....24....1T}, and the actual processing of the data will need to reflect this property. 
However, for the purpose of this study, our simplifying assumption has no significant impact on the results.

The likelihood function, Eq.~(\ref{eq:likelihood}), is expressed in terms of inner products $\langle . | . \rangle$ between time series. 
For two  time series, $a(t)$ and $b(t)$, whose Fourier transforms computed at the discrete frequency $f_k$ are ${\tilde a}(f_k)$ and ${\tilde b}(f_k)$, respectively, we define the inner product as
\begin{equation}
    \langle a | b \rangle = 2\sum_k \frac{{\tilde a}(f_k) {\tilde b}^*(f_k) + {\tilde a}^*(f_k) {\tilde b}(f_k)}{S(f_k)} \Delta f,
     \label{eq:inn_prod}
\end{equation}
where $S(f_k)$ is the noise PSD at frequency $f_k$, and $\Delta f = 1/T_\mathrm{obs}$ is the width of the frequency resolution bin for an observation time $T_\mathrm{obs}$. 

Two contributions enter the overall level of noise described by the PSD $S(f)$: the instrumental noise and the \emph{confusion noise}, due to background radiation from unresolved short-period binaries, primarily Galactic double WDs.
In our analysis we use the LISA instrumental noise model given in the LISA Science Requirement Document~\citep[SciRD, see][]{2021arXiv210801167B}. 
The Galactic binary confusion noise is modeled according to the estimate provided in the same document, see Section~9 of the SciRD. This estimate is based on assumptions about the performance of the LISA processing pipelines during science operations to resolve Galactic double WDs, and a specific astrophysical population model of these sources. Different assumptions about either of these two elements lead to slightly different levels of confusion noise, \citep[see \textit{e.g.}][]{2021PhRvD.104d3019K}. 
Furthermore, the Galactic population of double WDs used in the SciRD is generated independently of the population of \DCOS that we consider in this work.
The combination of these factors implies that during the actual observations the signal-to-noise ratio ($\mathrm{SNR}$)---and as a consequence the statistical uncertainty on the sources parameter measurements---may slightly differ from what we estimate here. 
Nonetheless, as the systems that will be identifiable as \DCOS have high $\mathrm{SNR}$ (see Section~\ref{sec:results}), our simplifying assumptions do not have any significant impact on the results presented here. 
In the future, however, a full self-consistent treatment of the \textit{entire} Galactic population of binaries will be necessary.

Under the assumptions described above, the optimal $\mathrm{SNR}$ at which a signal is observed can be simply expressed in terms of the inner product between the TDI time series:
\begin{equation}
    \mathrm{SNR} = \left[\sum_\alpha\langle y_\alpha|y_\alpha \rangle \right]^{1/2}\,.
    \label{eq:snr}
\end{equation}

The analysis that we carry out is based on a Bayesian inference code that some of the authors have been developing to prototype data processing approaches for LISA~\citep[\textit{e.g.}][]{2020ApJ...894L..15R, 2022arXiv220403423K, 2023MNRAS.522.5358F, 2024MNRAS.531.2817M, 2021PhRvD.104d4065B, 2023PhRvD.107l3026P, 2023PhRvD.108l4045P}.

\section{Astrophysical population}
\label{sec:astro_model}

In this study, we restrict ourselves to the canonical isolated binary evolution channel~\citep{2001A&A...375..890N,
2018MNRAS.480.2704L,
2020MNRAS.492.3061L,
2020MNRAS.494L..75S, 
2020ApJ...892L...9A, 
2020ApJ...898...71B,
2022ApJ...937..118W,
2024MNRAS.534.1707T}
in order to consider a plausible astrophysical Galactic population of binaries from which we can draw systems to simulate LISA data and observations. 
Other formation mechanisms are possible~\citep[\textit{e.g.}][]{
2018MNRAS.481.4775D, 
2018PhRvL.120s1103K,
2019PhRvD.100d3010S,
2025arXiv250118682X}, 
but our aim is not to provide a comprehensive assessment of LISA's performance, but rather to explore the potential of the mission to carry out mass measurements of \DCOS.

Specifically, we consider the model described in~\citet[][W22 hereafter]{2022ApJ...937..118W}, which in turn is based on the population synthesis model and software implementation by~\cite{2021MNRAS.508.5028B, 2022MNRAS.516.5737B}. 
We refer the reader to these papers for an extensive discussion of the astrophysical assumptions. 
W22 considers $20$ physical variations (see Sections 2.1--2.2., Table A1 and Appendix A1 of W22) of physical processes that affect the properties of the Galactic compact object binary population such as mass transfer, common-envelope evolution, supernova kicks and wind mass-loss physics, with the aim of exploring the LISA detection rate of double compact objects based on different assumptions.

\begin{figure}
    \centering
    \includegraphics[width=0.5\textwidth]{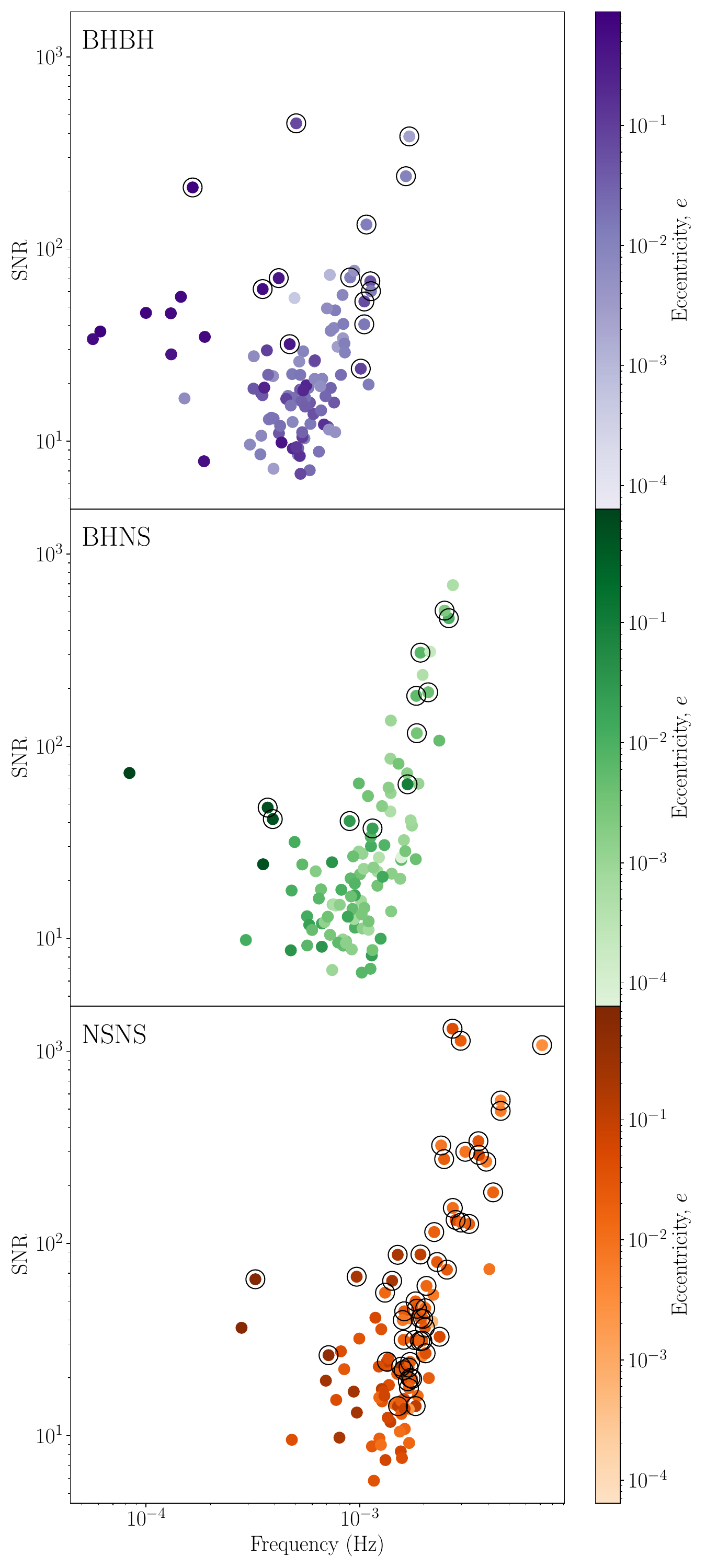}
    \caption{\label{fig:injFreqSNR}
    The frequency, $f_\mathrm{GW}$, see Eq.~(\ref{eq:fgw}), SNR, and eccentricity $e$ of the $300$ simulated systems used in this investigation ($100$ simulations of each source type). 
    The three source types BHBH, BHNS, and NSBH are shown in the top, middle, and bottom panel, respectively. 
    Each point represents a simulated signal's SNR and frequency. 
    The colour scale indicates the eccentricity. 
    The circled systems are those whose analysis meet the measurement criteria for ${\dot f}_\mathrm{GW}$, $f_\mathrm{PP}$, and $e$ described in Section~\ref{sec:analysis_measureability}.
    } 
\end{figure}

In this paper we use W22's \emph{fiducial model} (see Section~A.3 of W22). 
For this model, W22 estimate that during the nominal mission lifetime LISA will detect $74^{+8}_{-9}$ BHBHs, $42^{+7}_{-6}$ BHNSs and $8^{+3}_{-3}$ NSNSs, where the values are the mean and the $68\%$ Poisson probability range derived from the simulations. 
Model variations can have a considerable effect on the formation rate of these objects, the properties of the population, and consequently the number (and properties) of binaries that LISA will observe. The number of LISA-detected \DCOS can be as high (low) as $154^{+12}_{-13}$ ($6^{+2}_{-3}$) for BHBHs,  $198^{+14}_{-14}$ ($2^{+2}_{-1}$) for BHNSs, and $35^{+6}_{-6}$ ($3^{+1}_{-2}$) for NSNSs. 
These results are broadly consistent with those provided by similar studies, and a detailed comparison is provided in Table~12 of W22. 
This is a further indication of the important role that LISA observations may play in providing new information about \DCOS' demographics, and the astrophysical processes involved in their formation and evolution, provided that nature of these objects can be identified and distinguished from the much more numerous double WDs.

\begin{figure}
    \centering
    \includegraphics[width=0.5\textwidth]{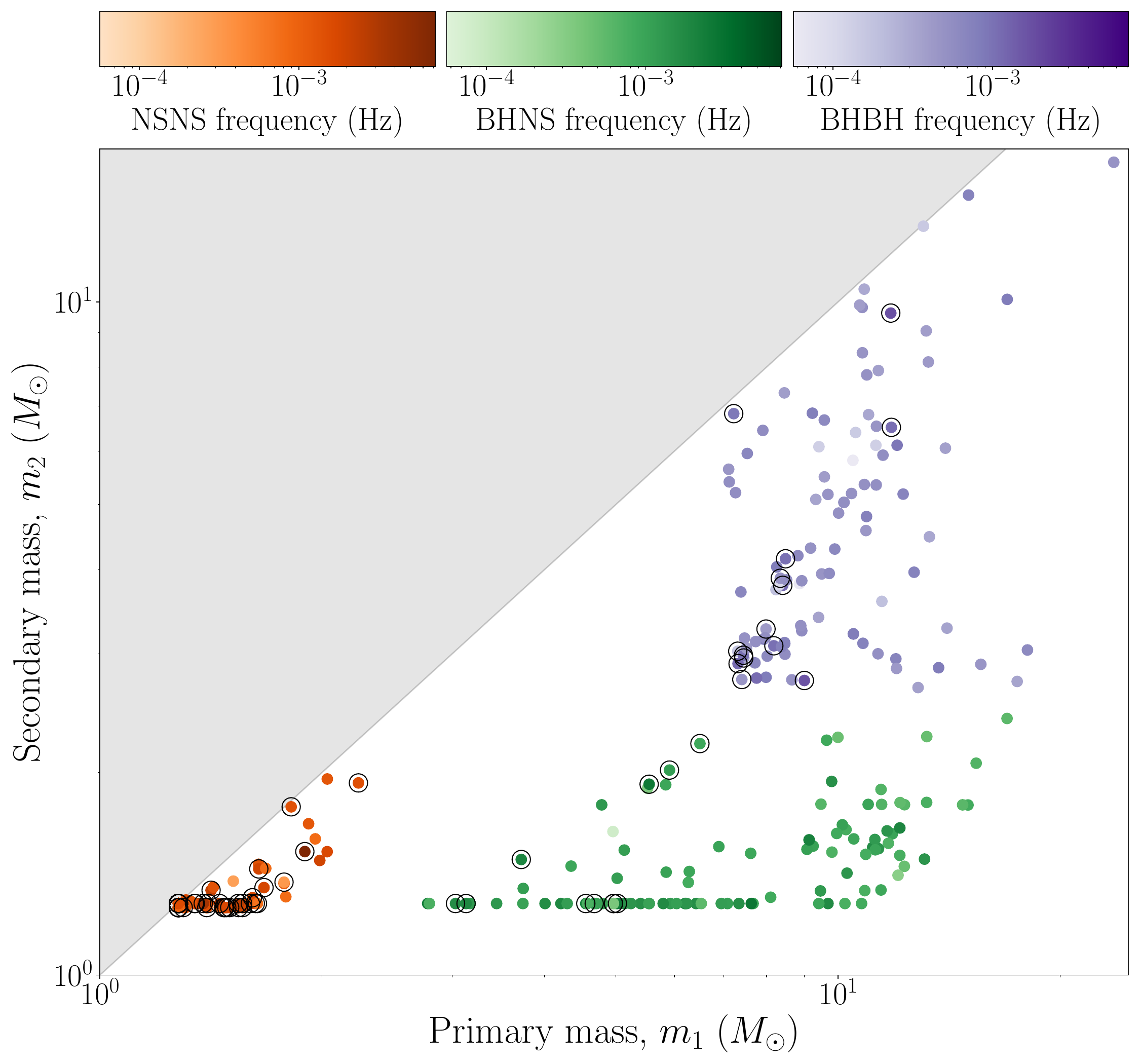}
    \caption{\label{fig:injm1m2_v2}
    The component masses ($m_1$, $m_2$) and frequency ($f_{\rm GW}$) of the $300$ simulated systems used in this investigation ($100$ simulations of each source type). 
    The three source types of BHBH, BHNS, and NSNS are shown in purple, green, and orange, respectively. 
    Each point represents a simulated system. 
    The colour scales indicate the frequency. 
    We use the convention $m_1>m_2$ and the grey shaded area indicates the region outside of this choice. 
    The circled systems are those whose analysis meet the measurement criteria for ${\dot f_{\rm GW}}$, $f_{\rm PP}$, and $e$ described in Section~\ref{sec:analysis_measureability}.
    }
\end{figure}

\begin{table*}[t!]
\begin{center}
\begin{tabularx}{.8\textwidth}{l||c|c|c||c|c|c||c}
\toprule
            & \multicolumn{7}{c}{Number of runs (/100) with measured parameters combinations} \\
Source type & \multicolumn{3}{c||}{Measured individually} 
            & \multicolumn{3}{c||}{Measured pairs} 
            & Measured all three \\
            & ${\dot f_{\rm GW}}$ & $f_{\rm pp}$ & $e$ & 
              ${\dot f_{\rm GW}}$ \& $f_{\rm pp}$ & ${\dot f_{\rm GW}}$ \& $e$ & $f_{\rm pp}$ \& $e$ & 
              ${\dot f_{\rm GW}}$ \& $f_{\rm pp}$ \& $e$ \\
\hline
BHBH & $29$ & $42$ & $34$ & $14$ & $13$ & $33$ & $13$ \\
BHNS & $47$ & $14$ & $13$ & $11$ & $10$ & $13$ & $10$ \\
NSNS & $60$ & $81$ & $71$ & $52$ & $49$ & $71$ & $49$ \\
\toprule
\end{tabularx}
\caption{\label{tab:measure}
Summary of the total number of simulations which meet the measurability criteria give by Eq.~(\ref{eq:measure}) for various parameter combinations.
There are $100$ simulations of each source type (BHBH, BHNS, NSNS).
The second, third, and fourth columns indicate the number of simulations for which ${\dot f_{\rm GW}}$, $f_{\rm pp}$, and $e$ are measured, respectively. 
The fifth, sixth, and seventh columns indicate the number of simulations for which two of the three parameters of interest are measured.
Finally, the eighth column indicates the number of simulations for which all three of the parameters are measured. 
These are the simulations shown by the dark-solid lines in Fig.~\ref{fig:inj_meas}.}
\end{center}
\end{table*}

\section{Results} 
\label{sec:results}

\subsection{Drawing binaries from the W22 distribution}

We use the distributions of the LISA-detected sources from the W22 fiducial model~\citep{W22-data} to draw the parameters of the binary systems that we then use to generate the TDI signal in LISA data.

W22 provide the distribution for sources detectable with a sky-averaged and inclination/polarisation-averaged $\mathrm{SNR} \ge 7$ for an observation time of $4\,\mathrm{yr}$. 
For each of the three categories (BHBH, BHNS, and NSNS) we draw $100$ sources from these distributions. 
Each source draw returns the values of $m_1$, $m_2$, $a$, $e$, and the 3D location of the source in the Galaxy. 
To uniquely define the LISA TDI observables, Eq.~(\ref{eq:y}), we need in addition to choose values for $\iota$, $\psi$, $\Phi_0$, and $\phi$. 
We draw these parameters from uniform distributions in the relevant range, as described in Table~\ref{tab:param}.
A summary of the main properties of the $300$ binaries considered here is provided in Figs.~\ref{fig:injFreqSNR} and~\ref{fig:injm1m2_v2}. 
Their locations in the sky are given in Fig.~\ref{fig:skymap}.

The distribution of detected sources provided by W22 is based on a LISA sensitivity curve that, in the meantime, has been (slightly) revised as described in Section~\ref{sec:LISAanalysis}. 
For this reason, for each source drawn from the W22 distribution, we first check that the binary produces a signal with $\mathrm{SNR} \gtrsim 6$ according to this more up-to-date noise model before including it into the analysis.
As the new LISA SciRD places no sensitivity requirement below $0.1\,\mathrm{mHz}$, we only consider binaries such that $f_\mathrm{GW} \ge 5\times 10^{-5}\,\mathrm{Hz}$. Finally, for computational reasons, we do not include any binary whose eccentricity is $\ge 0.9$. 
Binaries with $e \ge 0.9$ are sufficiently rare in the frequency band of interest (therefore our choice does not affect in any statistically significant way the sample that we consider), and such large values of $e$ require the inclusion of hundreds of harmonics in the signal to fully capture the $\mathrm{SNR}$, which significantly increases the computational burden of the analysis. For the actual analysis of the LISA data, efficient processing schemes will need to be developed to deal with the possible presence of short-period binaries on highly eccentric orbits, which could be fairly common in dense stellar environments.

\subsection{Analysis}

For each source drawn from the W22 distributions, we generate a dataset and process it according to the method described in Section~\ref{sec:LISAanalysis}. 
To keep the computational burden at a manageable level, we always limit the number of harmonics that we include in the signal to $n = 10$, see Eqs.~(\ref{eq:h+}) and~(\ref{eq:hx}). 
The likelihood function is constructed under the same hypothesis, which means that the signal model is identical for data generation and analysis. 
However, we caution the reader that, in the processing of the actual data from mission the number of harmonics to be included in the analysis will likely need to be larger in order to  capture the full SNR for systems with $e \gtrsim 0.1$, as the radiated GW power shifts progressively to higher harmonics, see \textit{e.g.}, Fig.~3 in \citet{1963PhRv..131..435P}.

We consider an observation time $T_\mathrm{obs} = 4\,\mathrm{yr}$ with no data gaps. 
We compute the posterior PDF on the model parameters, Eq.~(\ref{eq:posterior}) using \emph{zero-noise} realisations, and assuming that the noise spectral density is known. 
The Bayesian analysis is carried out on the 11-dimensional parameter space of the \textit{observable} parameters, Eq.~(\ref{eq:theta}), using priors that are reported in Table~\ref{tab:param}. 
For each of the $300$ binaries we compute the posterior PDFs on these parameters, which are the starting point of further analyses that we describe below.
The sampling is performed using \texttt{nessai}~\citep{nessai, Williams:2023ppp, Williams:2021qyt}, a nested sampling implementation~\citep{Skilling:2004}.

We are interested in whether specific parameters [within the whole set of $11$ that fully describe the signal, see Eq.~(\ref{eq:theta})] can be measured. 
In order to select the subset of events of interest, we adopt a practical and simple \textit{measurability criterion} that is based on the median and width of the one-dimensional marginalised posterior PDF of a given parameter $\theta_k$. We define
\begin{equation}
\mathcal{S}_k = \frac{\rm{med}(\theta_k)}{3\sigma_{\theta_k}}\,,
\label{eq:measure}
\end{equation}
where $\mathrm{med}(\theta_k)$ stands for the median of the distribution of a parameter $\theta_k$ and $\sigma_{\theta_k}$ is the difference between the $50{\rm th}$ and $16{\rm th}$ percentiles for the same parameter. If $\mathcal{S}_k>1$, we consider the parameter to be measured.

\begin{figure*}[htb!]
\begin{center}
\includegraphics[width=\textwidth]{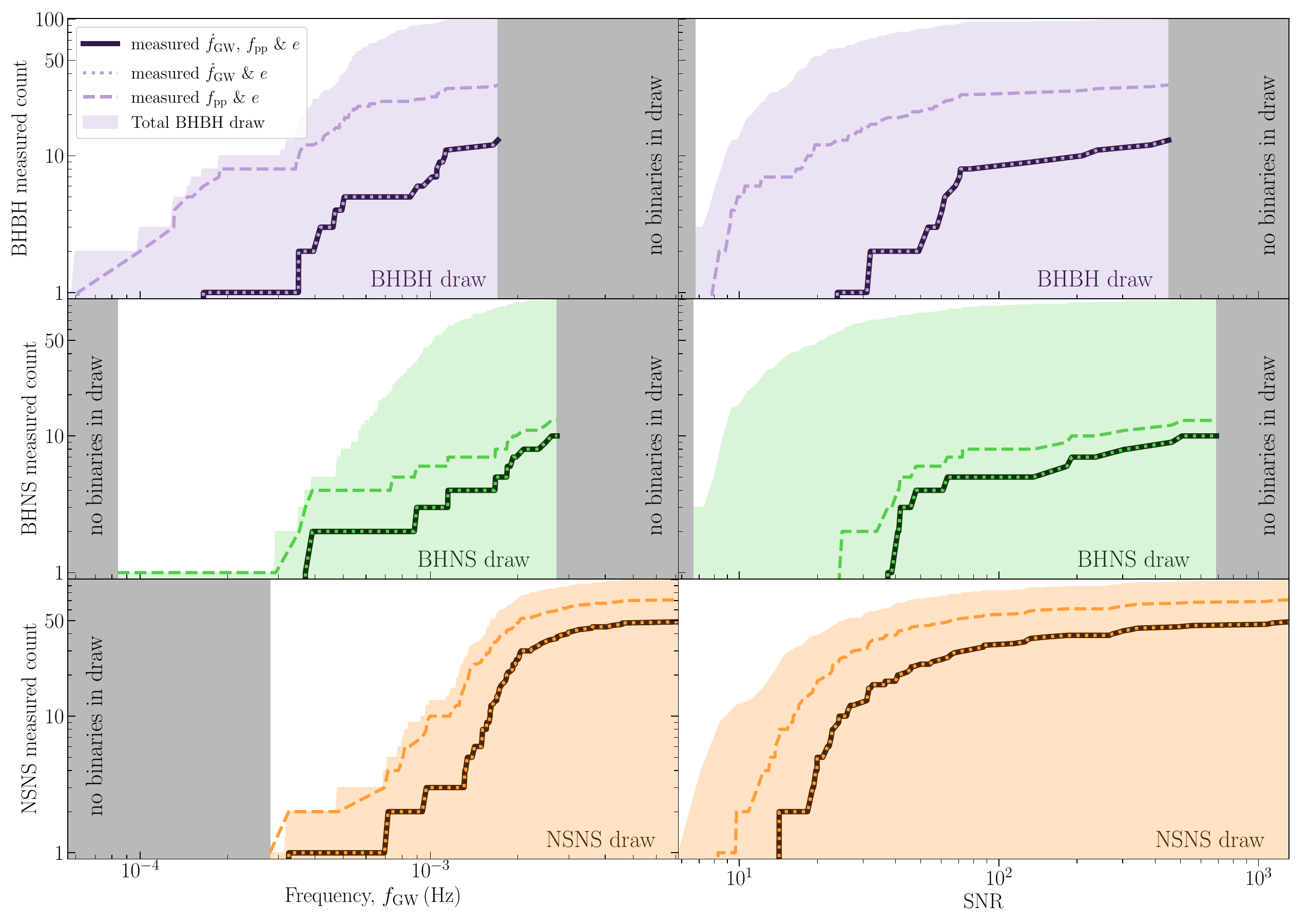}
\caption{\label{fig:inj_meas}
The cumulative number of binaries for which different combinations of ${\dot f}_\mathrm{GW}$, $f_{\rm PP}$, and $e$ are measured, according to the criteria defined in Eq.~(\ref{eq:measure}). 
The top, middle, and bottom rows show the three source types of BHBH, BHNS, and NSNS, respectively. 
The left and right panels show sources ordered by frequency and SNR, respectively. 
Within each panel, the shaded colouring shows the total cumulative count of injections for each source type ($100$ of each). 
The dotted and dashed lines show the cumulative count of injections with measured ${\dot f}_\mathrm{GW}$, $e$, and  measured $f_{\rm PP}, e$, respectively. 
The dark solid lines show the cumulative count of injections for which all three of  ${\dot f}_\mathrm{GW}$, $f_{\rm PP}$, $e$ are measured. 
A summary is given in Table~\ref{tab:measure}. 
The grey-shaded areas show regions where there were no binaries in our draw. 
}
\end{center}
\end{figure*}

\subsection{Individual mass measurements}
\label{sec:analysis_measureability}

As we have discussed in Section~\ref{sec:intro}, LISA will be able to measure the masses of a binary emitting a quasi-monochromatic signal if $f_\mathrm{GW}$, $\dot{f}_\mathrm{GW}$, ${f}_\mathrm{PP}$ and $e$ can be measured. 
In fact, from these quantities one can straightforwardly derive $M$ and ${\cal M}$, see Eqs.~(\ref{eq:fgw_dot}) and~(\ref{eq:fpp}), hence the mass of each of the binary's components. 
If a binary is detected, its frequency is always measured to $\approx\mathrm{nanoHertz}$ precision, and we will therefore restrict our discussion to the measurement of the three parameters $\dot{f}_\mathrm{GW}$, ${f}_\mathrm{PP}$, and $e$.

\begin{figure*}[htb!]
    \centering
    \includegraphics[width=0.24\linewidth]{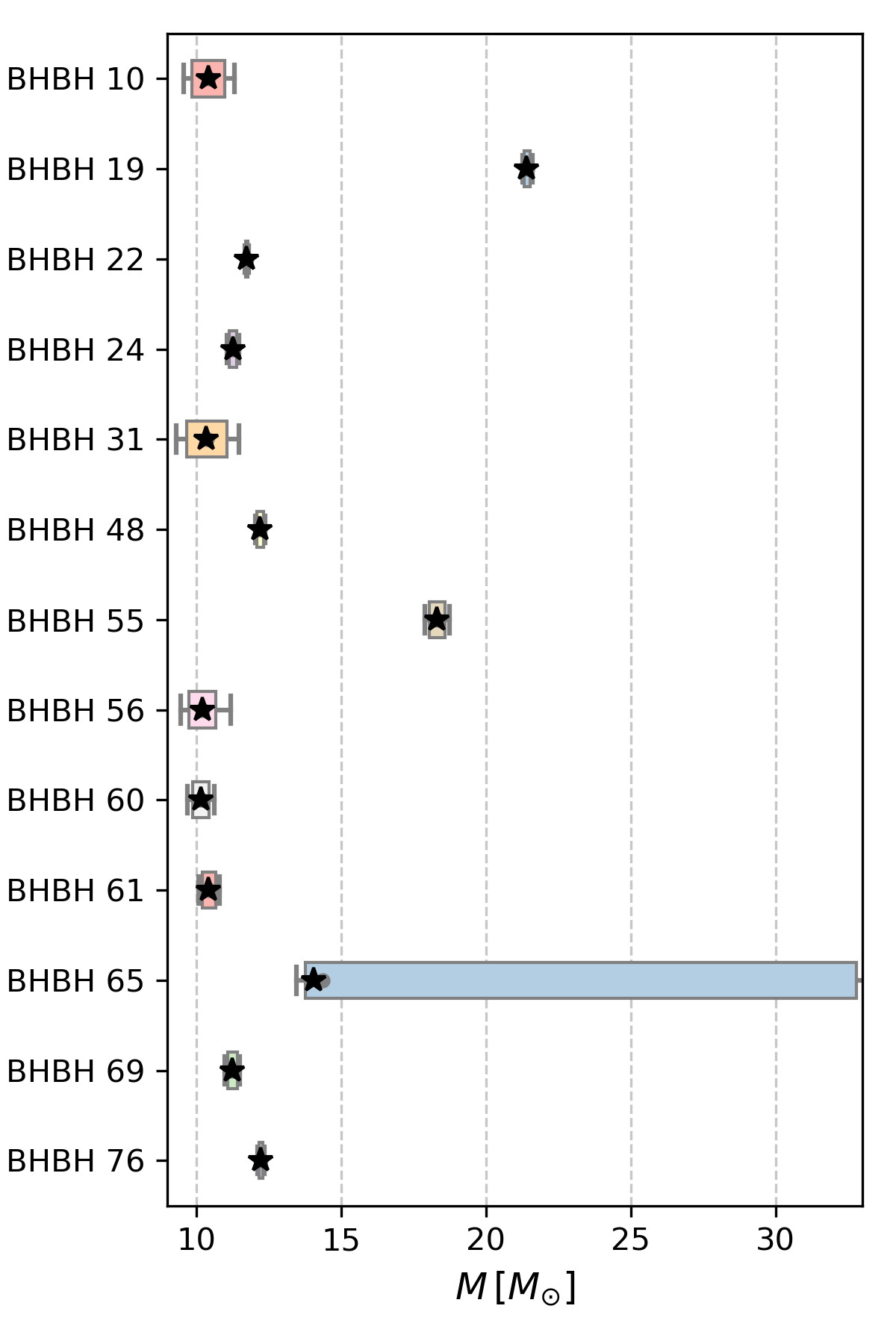}
    \includegraphics[width=0.24\linewidth]{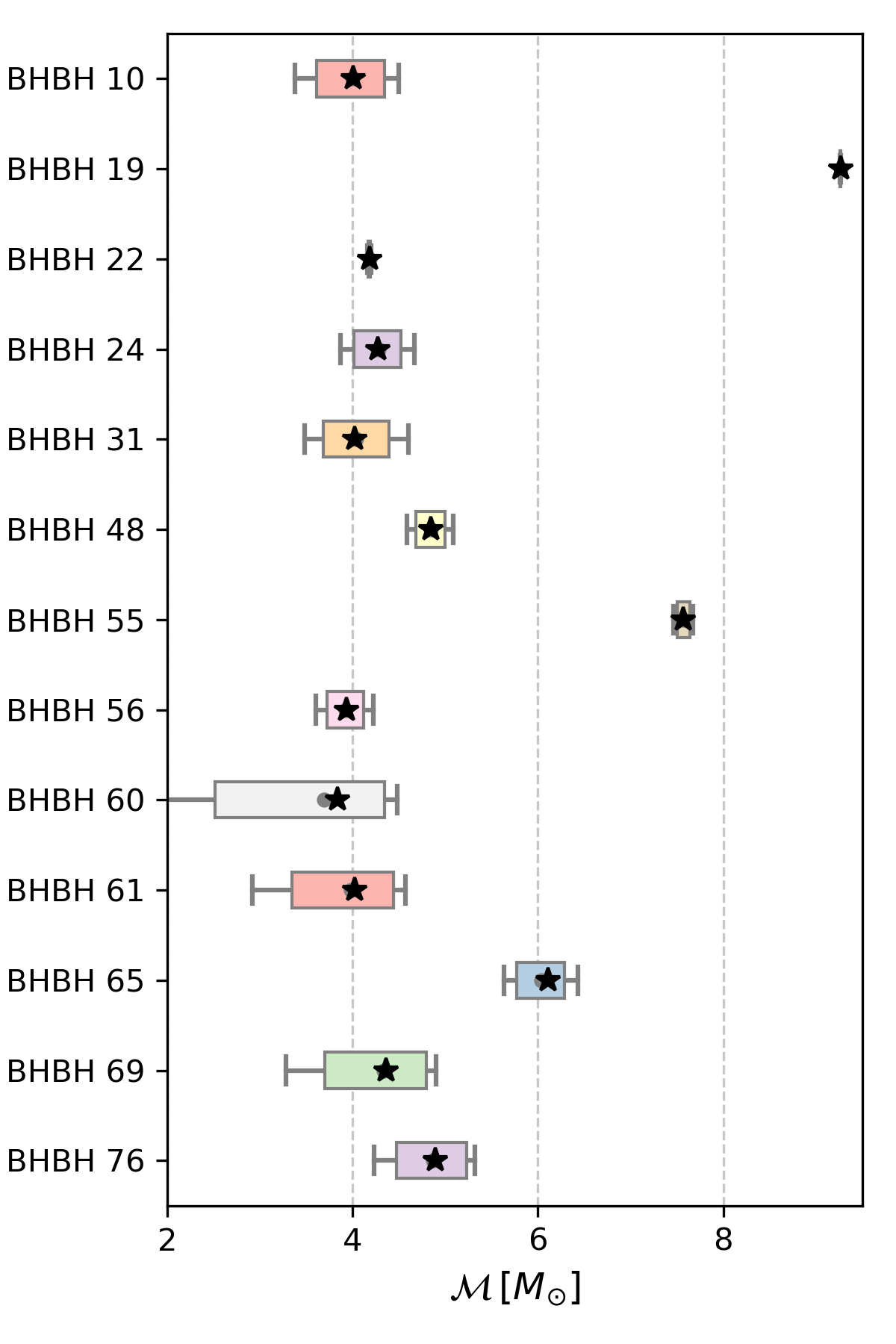}
    \includegraphics[width=0.24\linewidth]{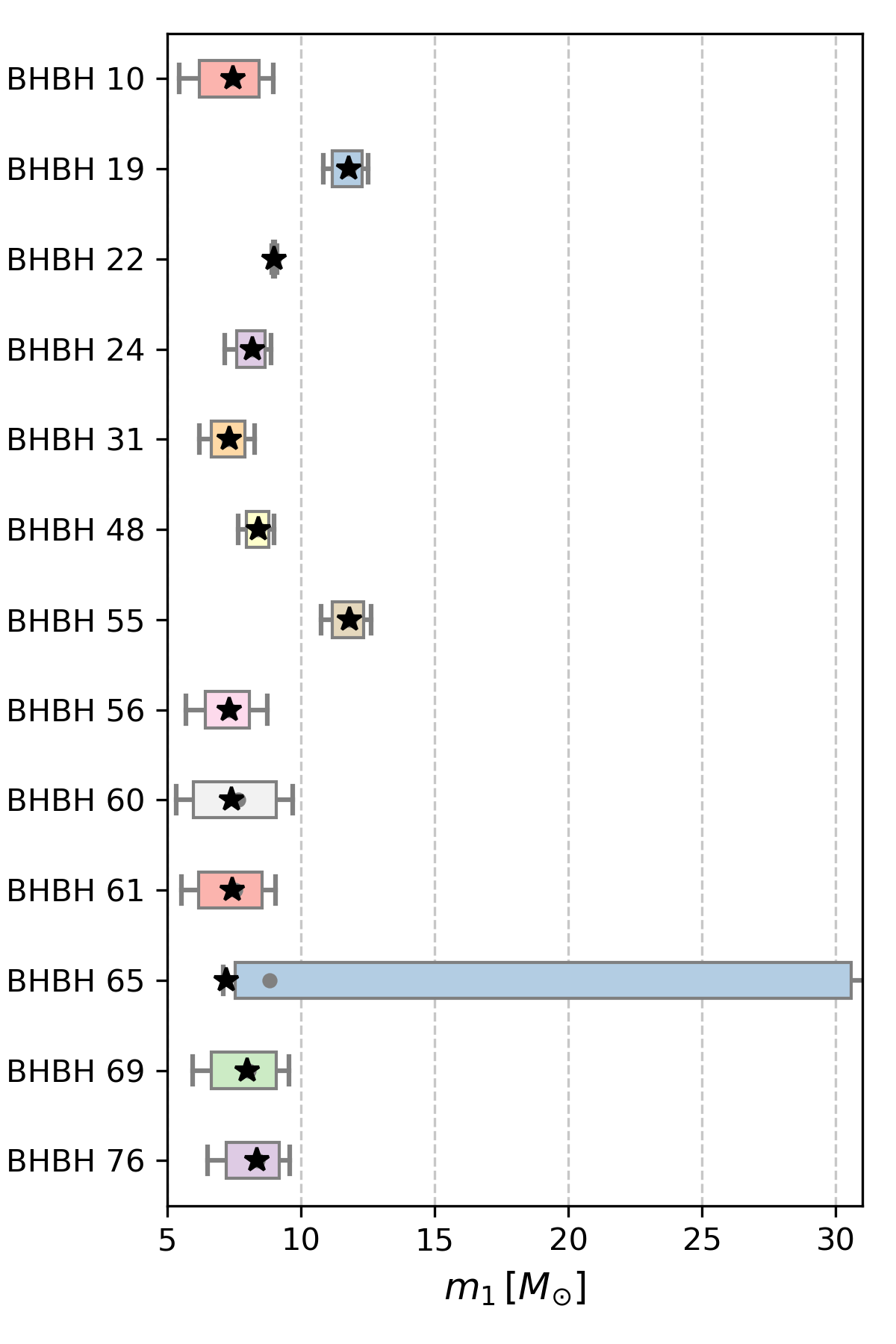} 
    \includegraphics[width=0.24\linewidth]{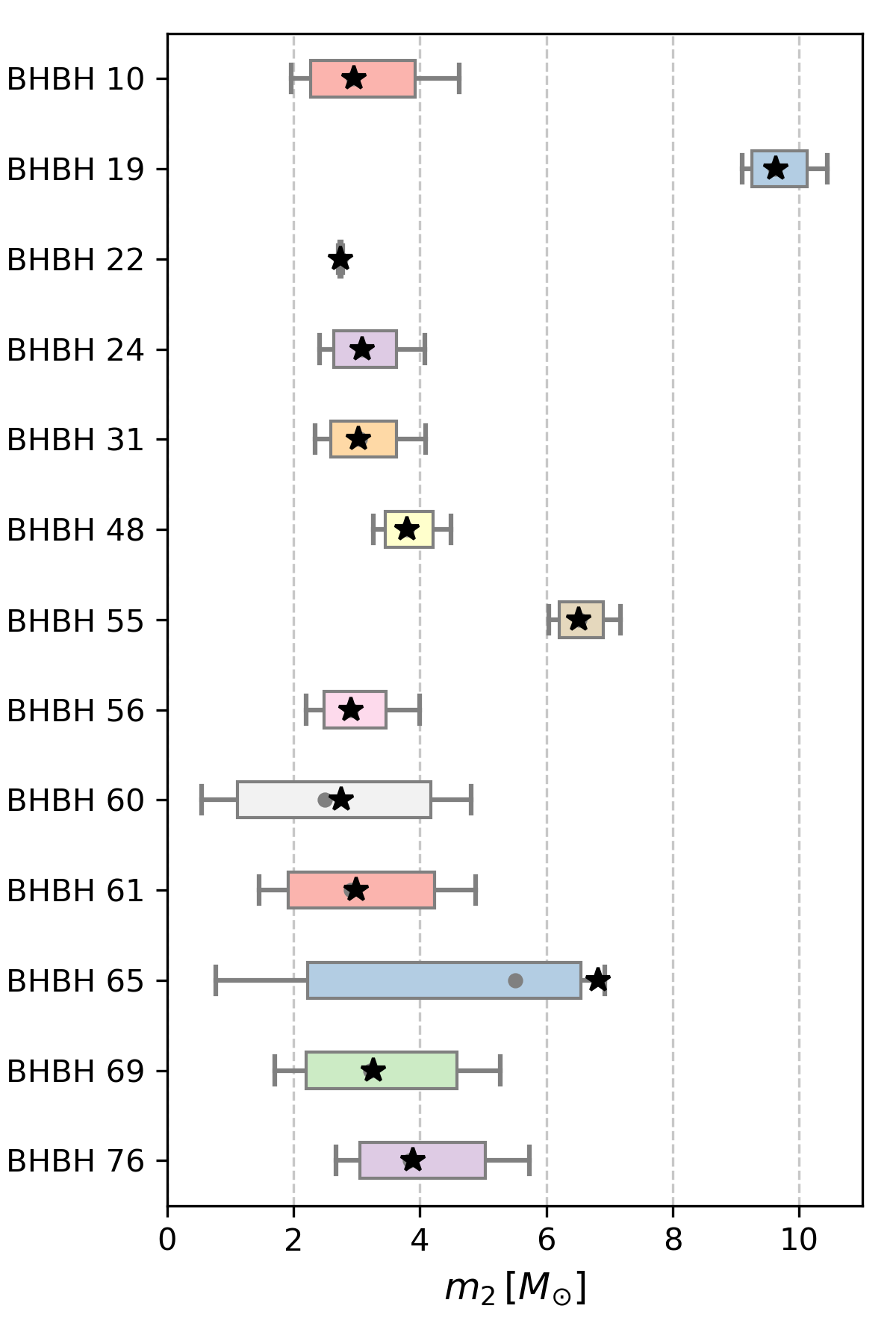}
    \vspace{-10pt}
    \includegraphics[width=0.24\linewidth]{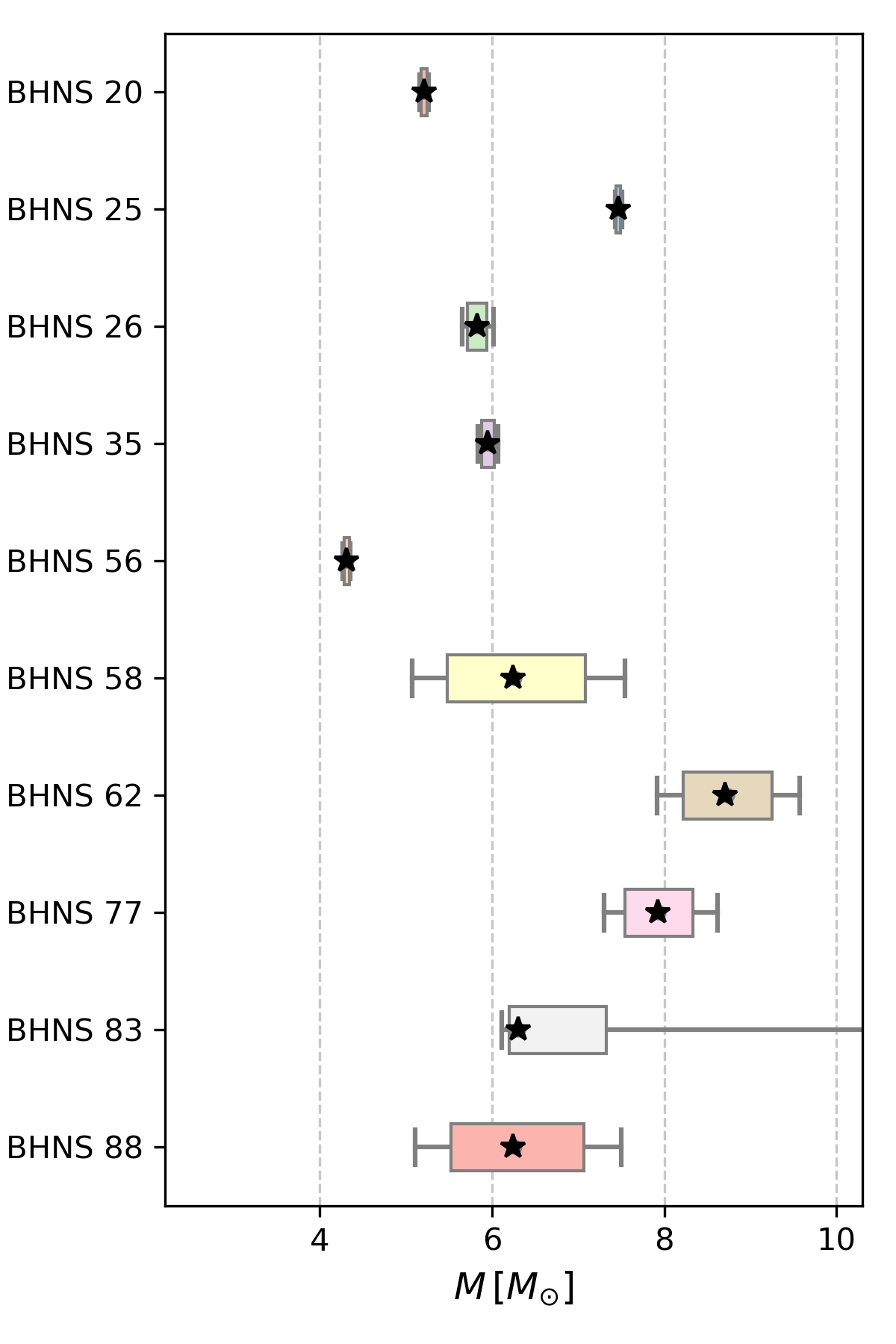}
    \includegraphics[width=0.24\linewidth]{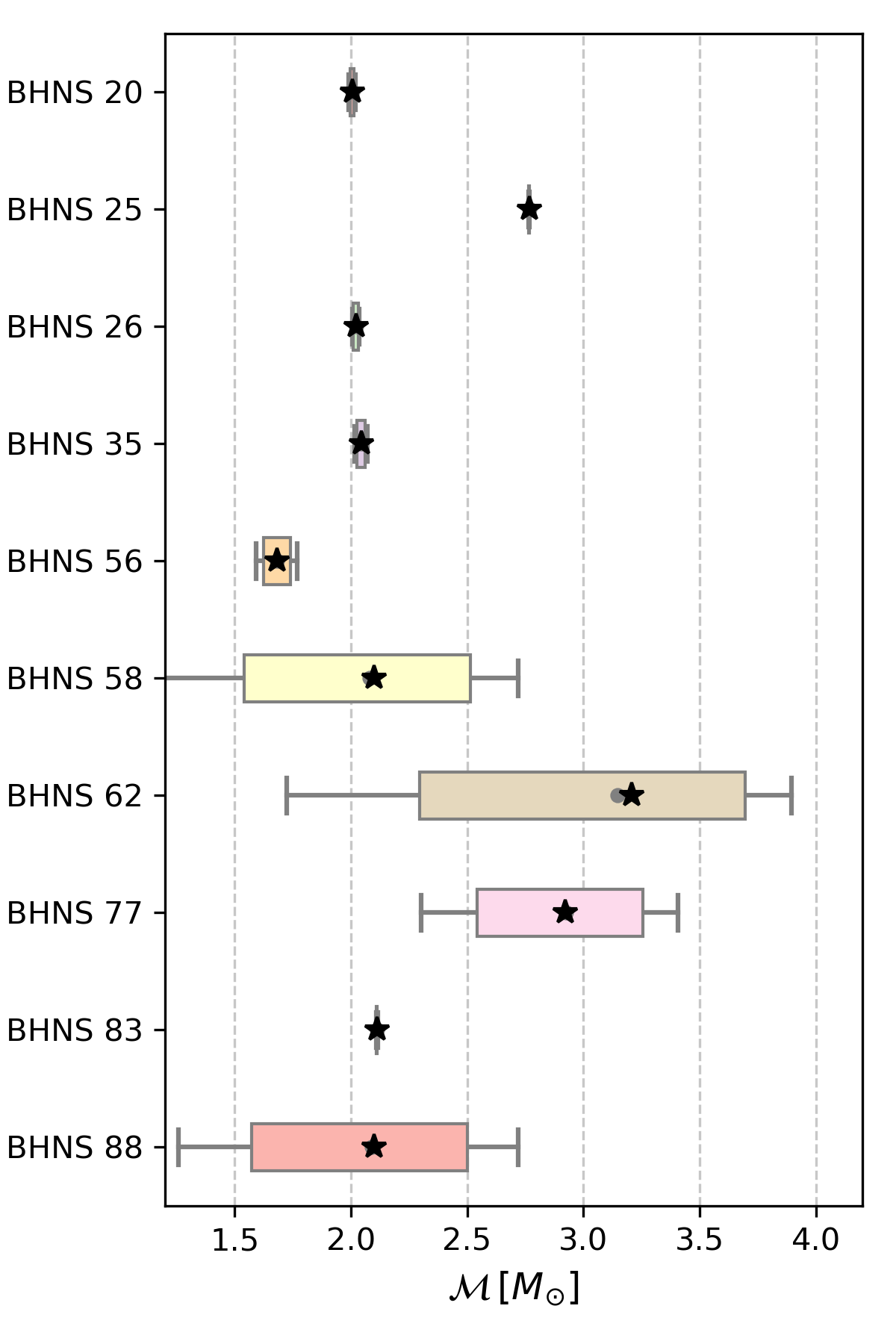}
    \includegraphics[width=0.24\linewidth]{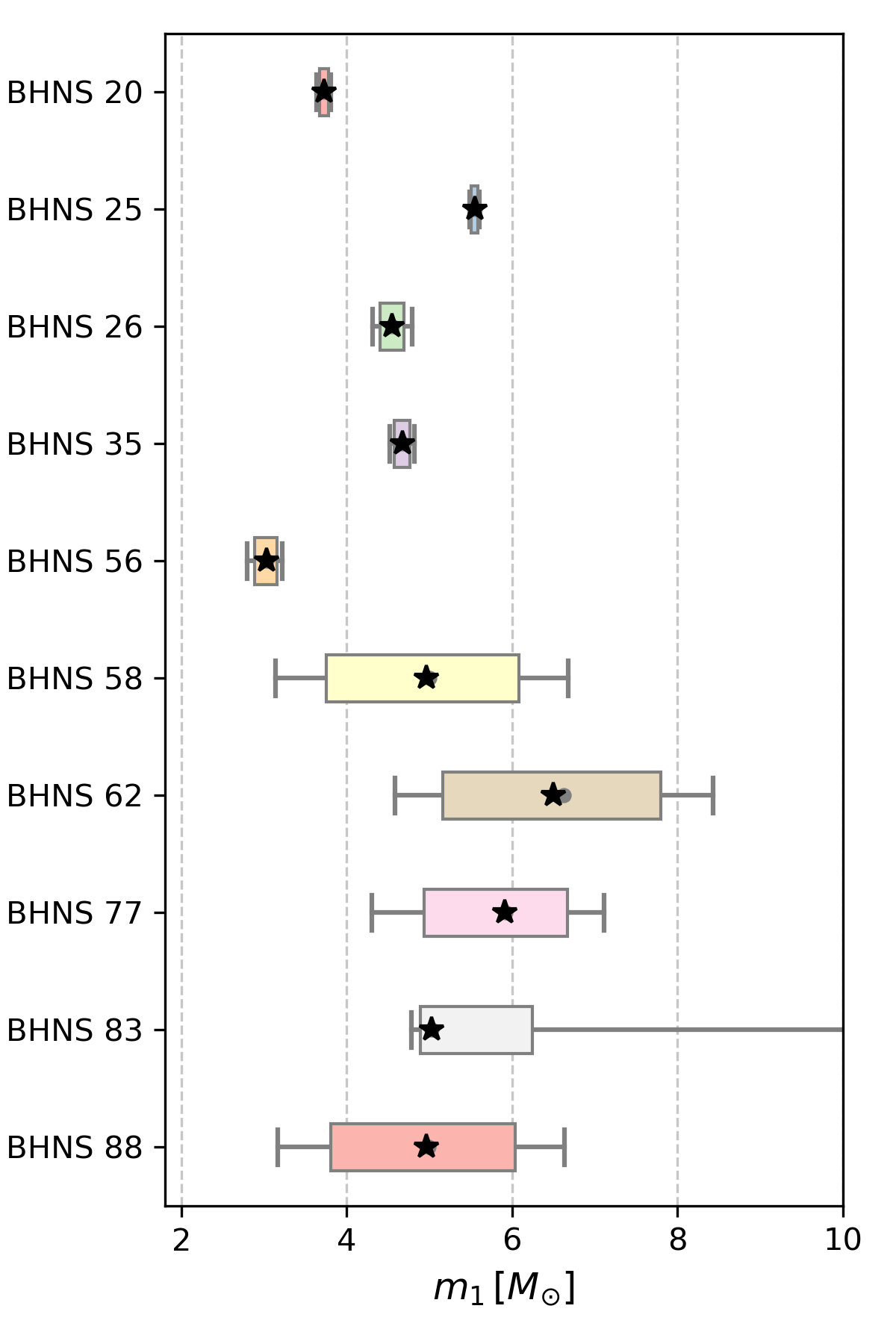} 
    \includegraphics[width=0.24\linewidth]{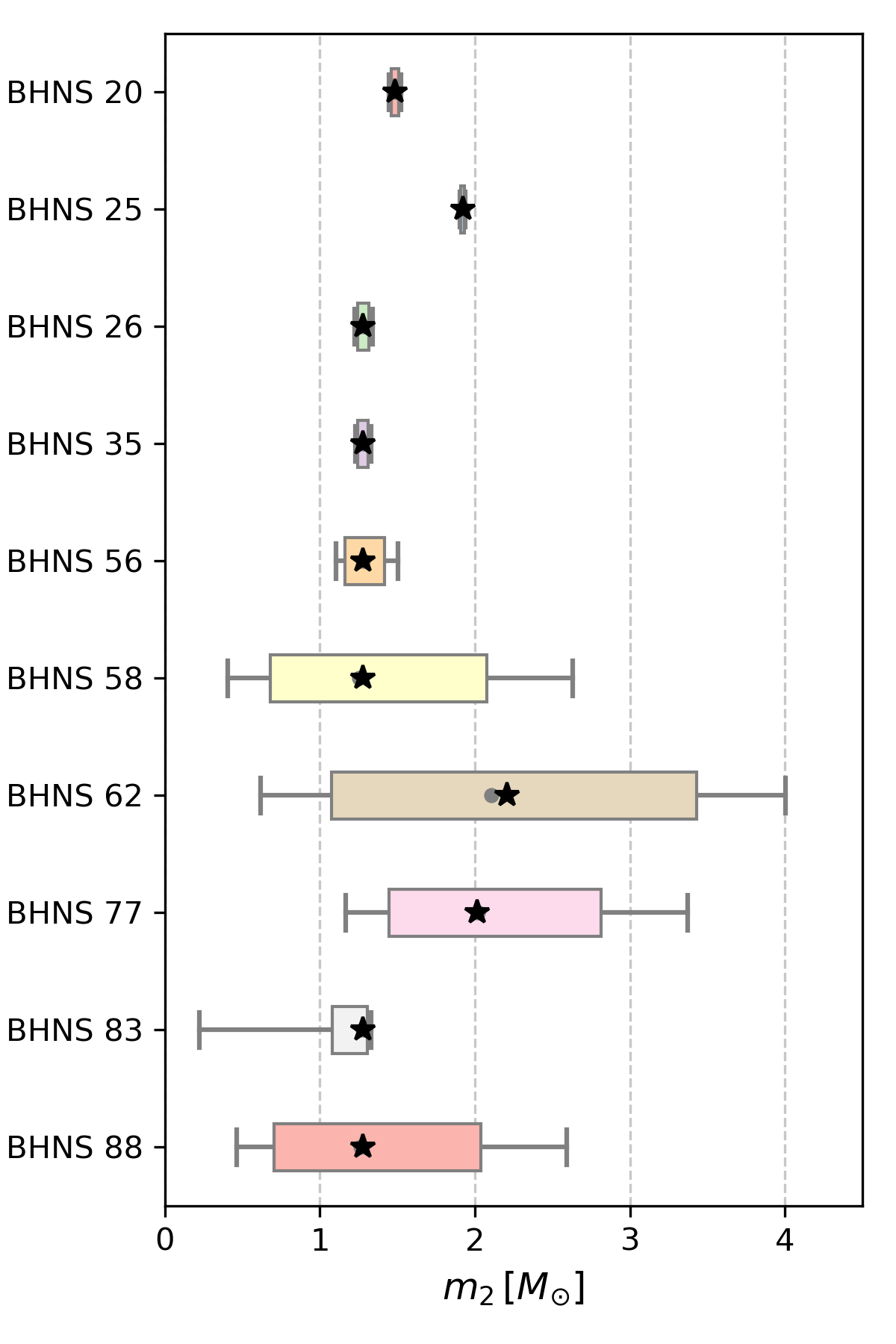}
    \vspace{-10pt} 
    \caption{Box plots showing the marginalised posterior distribution functions of selected parameters---total mass $M$, chirp mass ${\cal M}$, primary ($m_1$) and secondary ($m_2$) mass---of the $13$ BHBHs (top row) and $10$ BHNSs (bottom row) for which mass parameters can be measured.
    See also Table~\ref{tab:results_full} in Appendix~\ref{app:tables_box} and the accompanying data release.
    On the vertical axis we show the injection ID number.  Coloured boxes show the symmetric $90\%$ probability interval and thin grey lines show the symmetric $99\%$ probability interval. Dots denote the median value of the PDF, and stars correspond to the value of the injection (in some cases the star and the dot are essentially indistinguishable as they overlap). 
    }
    \label{fig:m_box_BHs}
\end{figure*}

\begin{figure*}[htb!]
    \centering
    \includegraphics[width=0.24\linewidth]{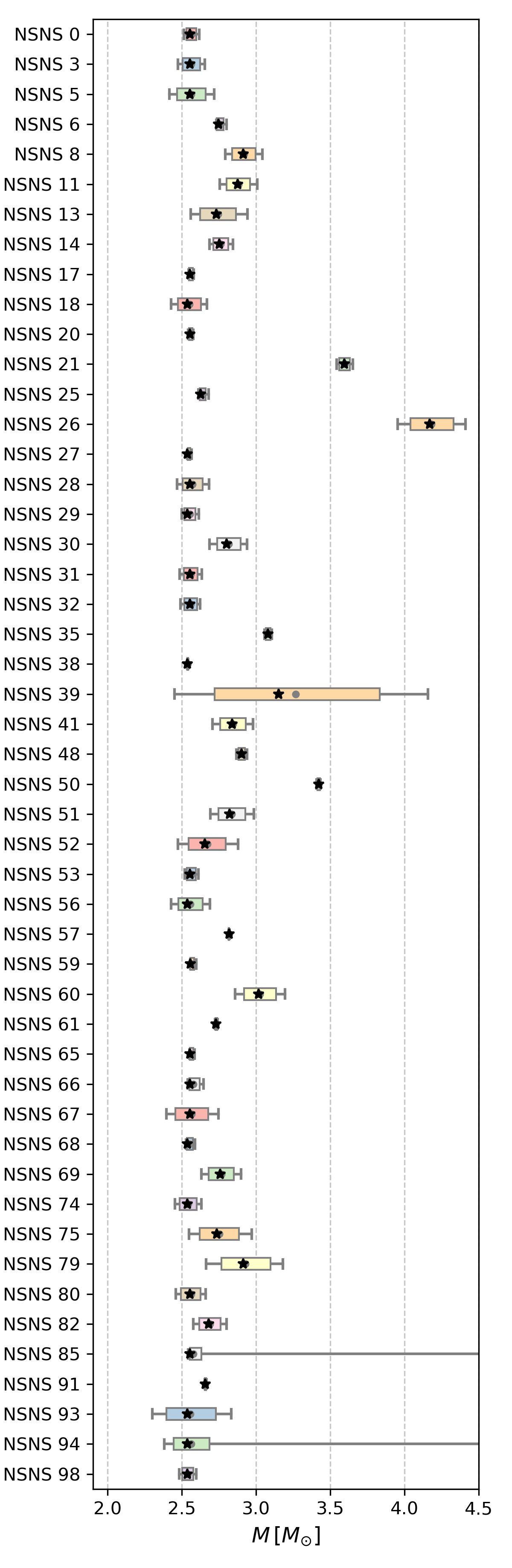}
    \includegraphics[width=0.24\linewidth]{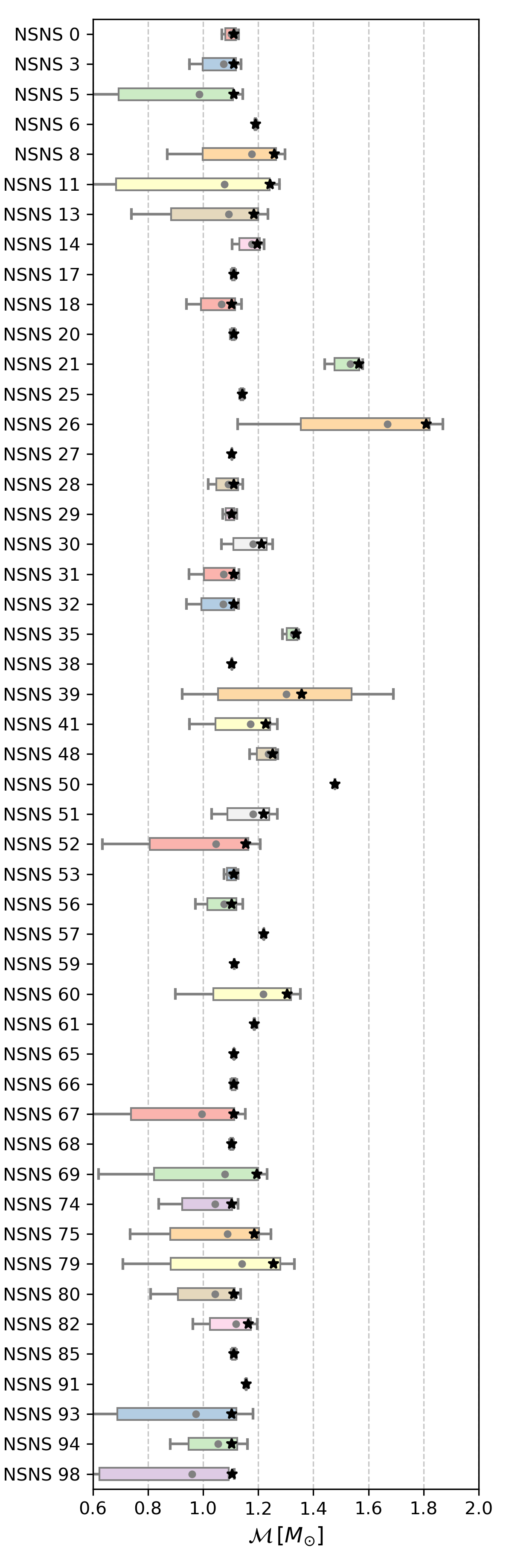}
    \includegraphics[width=0.24\linewidth]{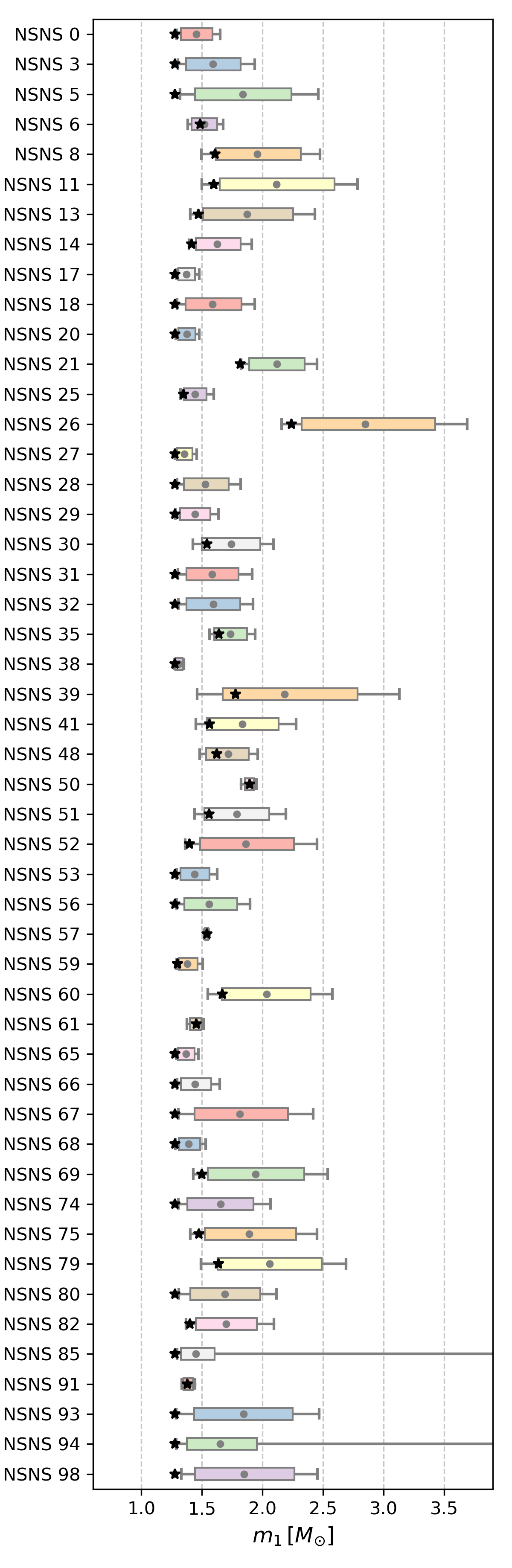} 
    \includegraphics[width=0.24\linewidth]{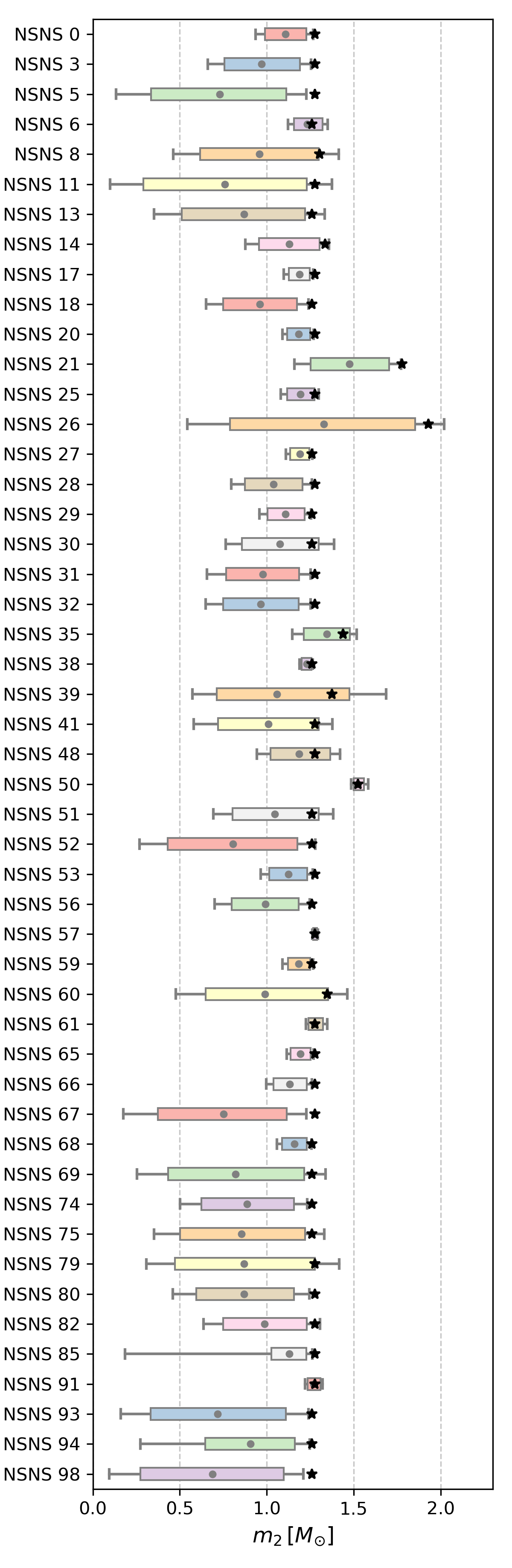}
    \vspace{-10pt} 
    \caption{Same as Fig.~\ref{fig:m_box_BHs} for the $49$ NSNSs for which mass parameters can be measured.
    See also Table~\ref{tab:results_full} in Appendix~\ref{app:tables_box} and the accompanying data release.
    }
    \label{fig:m_box_BNSs}
\end{figure*}

We apply the measurability criterion, Eq.~(\ref{eq:measure}), to $\dot{f}_\mathrm{GW}$, ${f}_\mathrm{PP}$, and $e$ for each of the $100$ signals from BHBHs, BHNSs, and NSNSs that we analyse. 
Fig.~\ref{fig:inj_meas} shows the cumulative number of systems, as a function of $f_\mathrm{GW}$ and $\mathrm{SNR}$, for which different combinations of these three parameters are measured, and a summary is given in Table~\ref{tab:measure}. 
The signal parameters and $\mathrm{SNRs}$ for the binaries for which $\dot{f}_\mathrm{GW}$, ${f}_\mathrm{PP}$, and $e$ are measured within the injection set considered here are given in Appendix~\ref{app:tables_box} in Table~\ref{tab:results_full} for selected systems and in the accompanying data release for all systems.

We find that all three parameters are measured for $13$ BHBHs, $10$ BHNSs, and $49$ NSNSs out of $100$ systems for each of the categories, see Table~\ref{tab:measure} and Fig.~\ref{fig:inj_meas}. 
These systems are circled in black in the scatterplots of Figs.~\ref{fig:injFreqSNR} and~\ref{fig:injm1m2_v2}. 
This implies that for double compact objects formed through the isolated binary evolution channel (consistent with W22), LISA will be able to measure the masses of $\approx 10\%$ of BHBHs and BHNSs and $\approx 50\%$ of the NSNSs. 
The main reason for the significantly larger fraction of NSNSs is that these binaries will be observed at higher frequencies and, as a consequence, higher SNR than BHBHs and BHNSs, see Fig.~\ref{fig:inj_meas}. 
We stress that the \textit{total} number of NSNSs that LISA is expected to detect is $\approx 5$--$10$ times smaller than BHBHs and BHNSs, hence the total number of binaries for which masses will be known is likely comparable in each of these categories.

The measurability of $\dot{f}_\mathrm{GW}$, ${f}_\mathrm{PP}$, and $e$ as a function of $\mathrm{SNR}$ and $f_\mathrm{GW}$ shows similar trends for the three classes of \DCOS. 
The number of systems for which the combination ($f_\mathrm{PP}, e$)---\textit{i.e.}, the total mass---can be measured is greater than those for which $(\dot{f}_\mathrm{GW}, e)$---\textit{i.e.}, the chirp mass---can be determined. 
The number of BHBHs (NSNSs) for which the total mass can be determined is $\approx 3 (2) $ times greater than those for which the chirp mass can be determined, whereas it is comparable for the BHNS systems.
We note that for all the categories, if the chirp mass can be measured, then also the total mass can be measured. The opposite is not true. For BHBHs (NSNSs) we find that the total mass can be measured for $\approx 30\%\,(70\%)$ of the systems.

From the joint posterior PDF on $f_\mathrm{GW}$, $\dot{f}_\mathrm{GW}$, ${f}_\mathrm{PP}$, and $e$, we can derive posterior distributions on any choice of two independent mass parameters. 
Here we consider the combination $({\cal M}, M)$, as ${\cal M}$ and $M$ are directly proportional to the parameters that describe the signal, $\dot{f}_\mathrm{GW}$ and ${f}_\mathrm{PP}$, respectively, and $(m_1, m_2)$, which are more natural parameters to connect observations to astrophysics. 
We note that in processing the data, the priors that we place on $\dot{f}_\mathrm{GW}$ and ${f}_\mathrm{PP}$ (see Table~\ref{tab:param}) are agnostic about the physical process driving the frequency evolution of a binary. 
At the postprocessing stage to derive the masses, we make the additional assumption that the binary evolves purely due to gravitational radiation reaction. 
There are therefore posterior samples that do not correspond to a physical system. For this reason, we impose additional prior cuts under the assumption that gravitational radiation reaction is the only physical effect at play. 
Specifically, the symmetric mass ratio, $\eta$ Eq.~(\ref{eq:eta}), is subject to the physical constraint $\eta \le 1/4$, and we disregard the samples that do not satisfy this condition.

A summary of the posterior distributions for the mass parameters is given in Fig.~\ref{fig:m_box_BHs} for BHBHs and BHNSs, and in Fig.~\ref{fig:m_box_BNSs} for NSNSs. 
The median and 90\% symmetric probability intervals of the posterior PDFs are given in Appendix \ref{app:tables_box} in Table~\ref{tab:results_full} (for selected systems) and in the accompanying data release (for the full set of systems).
As examples, two-dimensional posterior distributions for selected systems are provided in the Figs.~\ref{fig:BHBH19}, ~\ref{fig:BHNS88}, and~\ref{fig:NSNS35}.

The fractional errors for ${\cal M}$ and $M$ range from a few $\times 10\%$ to sub-$1\%$ level. 
The fractional errors on the \textit{individual} masses, $m_{1,2}$ range from $\approx 100\%$ to $\approx 1\%$. 
The measurement error depends on the actual value of the frequency, eccentricity, and masses of a binary, and the distance to the source, which affects the SNR at which a binary is detected. 

For the sample of BHBHs considered here, we see that for essentially all the systems for which one can measure their mass one would be able to infer with a high degree of confidence that the primary (more massive) component of the system is a BH. 
For the secondary (less massive) component, the conclusion is less obvious and depends on how close is the mass of the object to the NS mass range. Based on causality arguments, the maximum mass of a NS is $\lesssim 3\,M_\odot$ \citep{1974PhRvL..32..324R, 1996ApJ...470L..61K}. 
The actual value of the maximum mass of a NS produced by astrophysical processes is currently constrained to the range $\approx 2.0-2.7\,M_\odot$~\citep{2016ARA&A..54..401O, 2020PhRvD.102f3006S, 2021ApJ...918L..28M, 2021ApJ...918L..29R, 2021PhRvC.104c2802L}, see also \citet[]{2024arXiv240711153C} and references therein for a recent review. 
For example, we see that one would clearly conclude that BHBH-19 and BHBH-55 are BHBHs.  For the others, the tail of the distribution of $m_2$ extends to sufficiently low values, which can reach $\approx 2\,M_\odot$, and, in some cases, even below $1\,M_\odot$.

\begin{figure*}[htb!]
\centering
\includegraphics[width=\textwidth,trim={0 2cm 0 0cm} ]{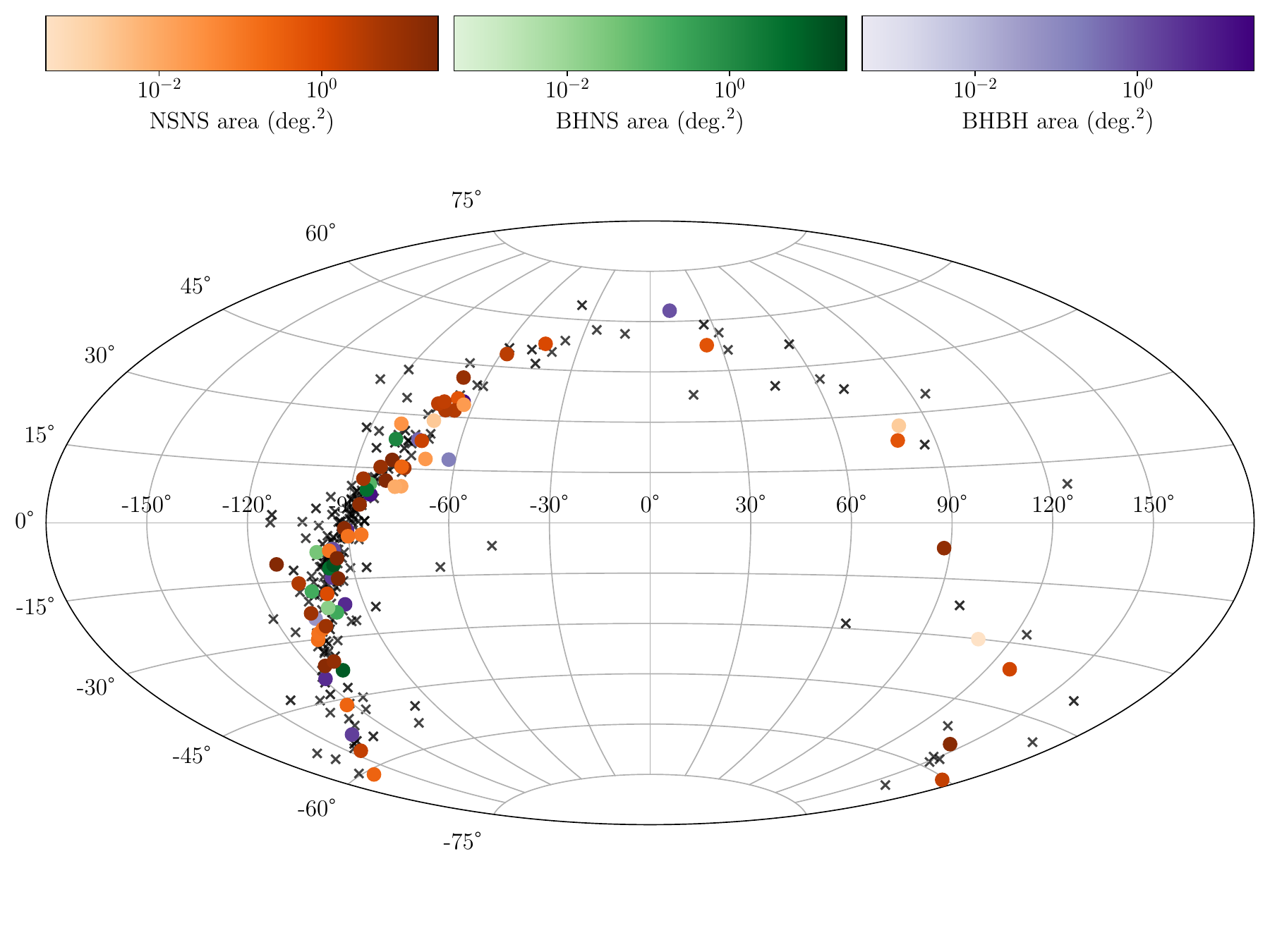}
\caption{\label{fig:skymap}
Sky map of injected BHBH, BHNS, and NSNS binaries in ecliptic coordinates. 
Each black cross marker represented an injected binary. 
The circles show the binaries for which we measure ${\dot f}$, $f_{\rm pp}$, and $e$. 
The BHBH, BHNS, and NSNS systems are shown in purple, green, and orange, respectively.
The color shading indicates the size of the recovered sky areas ($90\%$ credible area).
}
\end{figure*}

A similar behaviour can be observed for BHNSs. 
We find that in essentially all the cases one can be confident that the primary component is a BH (an exception is BHNS-56 whose primary has a mass of $3\,M_\odot$). 
In most cases the secondary mass is sufficiently well constrained that one can reasonably place it in the NS mass range, but there are exceptions, \textit{e.g.}, BHNS-62 and BHNS-77. 
The NS mass range also overlaps with the one allowed for WDs. The Chandrasekhar limit is $1.44\,M_\odot$, and the most massive WD known to date has a mass of $1.346\pm 0.019\,M_\odot$~\citep{2021Natur.595...39C}. 
In several circumstances mass measurements will be consistent with the secondary object being a NS or a WD close to the Chandrasekhar limit. 
To what extent LISA will be able to discriminate between NSs and WDs simply based on mass measurements is outside the scope of this paper. 
However, we note that a WD orbiting a BH in a $\approx 10^3\,\mathrm{s}$ orbit will be affected by mass transfer and tidal effects causing a departure from the pure radiation reaction evolution, which may provide additional clues to identify the nature of the object.

For NSNSs, we see that for essentially all the systems considered here the mass measurements are sufficiently precise to exclude that the binary contains a BH. 
However, LISA's ability to unambiguously identify the binary components as NSs is affected by the same issue that we have discussed above, \textit{i.e.}, the overlap of the NS and WD mass ranges. 
An additional piece of information that could be valuable is that the eccentricity of the binary will be known, and different from zero. 
Figure~\ref{fig:de_box_BNS} in Appendix~\ref{app:tables_box} illustrates that eccentricity can be measured with a fractional error $\approx 10\%$ and the smallest eccentricity of NSNSs in our sample is $0.0031$ (selected results for the recovered eccentricity are included in Table~\ref{tab:results_full} in Appendix~\ref{app:tables_box} and full results in the data release). 
The orbits of double WDs are expected to be circular to a high degree by the time they enter the LISA band, see \textit{e.g.} \cite{2024arXiv240207571C} and references therein. 
This should avoid incorrectly identifying a NSNS as a double WDs. 
However, a small number of double WDs with a residual eccentricity may be present in dense environments, such as globular clusters, and LISA could detect up to $\sim 10$ eccentric double WDs~\citep[see Fig. 5 in][]{2025A&A...702A.131H}. 
This number is roughly comparable with the total number of NSNSs observed by LISA during its nominal mission lifetime. 
Furthermore, distinguishing a NSNS from a neutron star-white dwarf may be equally challenging. 
LISA is expected to observe ${\cal O}(100)$ neutron star-white dwarf binaries with non-negligible eccentricity. 
Of these, it is expected that eccentricity and at least chip mass will be measureable for $\gtrsim 10$ of systems~\citep[see Table 1 in][]{2024MNRAS.530..844K}. 
These systems are in general ``heavy'' with a chirp mass $\approx 1\,M_\odot$ which is comparable to the one of a typical NSNS~\citep[chirp mass measurements can also provide clues to formation history; see, e.g.,][]{2026MNRAS.546ag117M}.

On purely observational grounds based on mass measurements, this uncertain classification is likely to be difficult to resolve. An open question that remains and deserves future investigation (beyond the scope of this paper) is to what extent the ambiguity can be broken. 
As we discuss in the next Section, the locations of these sources in the Milky Way will be known, which enables follow-up observations with other telescopes. 
They could yield the identification of an optical counterpart, which would immediately resolve the inconclusive classification provided by LISA. 
Follow-up observations would also provide information about the environment in which a binary resides, which may offer further clues about the likely nature of the system.

In summary, the results presented above are encouraging for the purpose of determining the nature of the constituent objects of a Galactic DCO. 
In the first instance, excluding the possibility that the binary is one of the $\sim 10^4$ double WDs could be a criteria to trigger more detailed investigations and targeted follow-up observational campaigns with other observing facilities on a reduced number of candidate targets.

\subsection{Locating the binaries in the Milky Way}

LISA will be able to determine the 3D locations in the Milky Way of all the binaries for which the system masses are measured. 
In fact, the motion of the LISA constellation during the year-long observation of these systems induces sky-location-dependent modulations in the phase and amplitude of the TDI outputs that allow for the determination of $(l, b)$ \citep{1998PhRvD..57.7089C}. 
In addition, determining $\dot{f}_\mathrm{GW}$ and $e$, which is necessary for mass measurements, breaks the ${\cal M}-D$ degeneracy in the measured amplitude of the GW signal, see Eq.~(\ref{eq:A}). 
The distance to the source can therefore be determined.

We find that the typical fractional errors on the distance are $\Delta D/ D \approx 1\%$--$100\%$ (see Appendix~\ref{app:tables_box} and the data release). 
The $90\%$ probability region of the projected location of the source in the sky is $\Delta\Omega \approx 10^{-2}$--$10\,\mathrm{deg}^2 $. 
We show a summary sky map of the systems considered here in Fig.~\ref{fig:skymap}.
The values for $\Delta\Omega$ for selected systems are given in Appendix~\ref{app:tables_box} and full results in the accompanying data release. 

The ability to place each of these binaries in a specific region of the Galaxy will be important to enable further studies. 
It will enable targeted observations with a slew of telescopes that could, \textit{e.g.}, rule out WDs as constituents of a binary and/or shed some further light on the nature of the compact objects. 
It will establish the Galactic (or otherwise) nature of a source, as well as providing additional information about the specific environment that harbours a binary and therefore clues about the context in which their progenitor stars have evolved.

\section{Conclusions}
\label{sec:concl}

We have considered LISAs' ability to measure the individual component masses of Galactic binaries composed of BHs and NSs. 
For a population of \DCOS consistent with the theoretical predictions by~\citet{2022ApJ...937..118W}, we have shown that for $\approx 10\%$ of the detected BHBHs and BHNSs, and $\approx 50\%$ of the detected NSNSs, respectively, LISA will be able to provide individual mass measurements, typically at the $10\%$ level. 
In addition, LISA will measure the orbital eccentricities of these systems and their locations in the Galaxy.

Based on the LISA detection estimates by~\citet{2022ApJ...937..118W}, this represents a small but precious sample of $\sim 10$ GW-selected Galactic \DCOS, which complements the results of other surveys in radio, X-ray, astrometry, and micro-lensing. 
It will play an important role in providing better understanding of the formation and evolution history of BHs and NSs and the environments in which they reside. 
These numbers are only indicative, and should be taken \textit{cum grano salis} as they are based on a (representative) theoretical population of binaries formed through the standard binary evolution channel. 
Other formation channels, such as those in dense environments, could play an important role in determining the number and properties of systems detected by LISA, and our study provide evidence for the rich information that LISA could provide.

Our analysis is based on forward modeling and does not directly address the processing challenges that will need to be tackled to actually make these measurements. 
This is outside the scope of this paper, and we will return to these aspects in the future. 
We will, however, highlight some of the assumptions that we have made and aspects that will need to be addressed in the future for practical implementations of analysis strategies.

We have assumed that the noise is known a priori, and is Gaussian and stationary throughout the mission. 
Although this assumption is unlikely to significantly affect the results of the work that we have presented here, it will need to be accounted for in the actual implementation of the processing of the LISA dataset, see \textit{e.g.},~\citet[]{2023PhRvD.107f3004L, 2023MNRAS.522.5358F, 2024PhRvD.110b4005S, 2025PhRvD.111b4060K, 2025PhRvD.111j3014D}. 
We have also ignored the issue of the presence of gaps~\citep{2025arXiv250217426B} and noise transients~\citep{2025arXiv250519870M} that effectively reduce the total observation time. 
As the signals from Galactic \DCOS are long lived, in multiyear observations the effect of a shorter effective observing duration is simply reflected in the measurement errors of each parameter as a larger statistical uncertainty, as the measurement errors scale as $1/\mathrm{SNR} \sim T_\mathrm{obs}^{-1/2}$. 
We have also ignored possible effects coming from source confusion. 
We have shown, however, that the systems for which mass measurements are going to be possible have sufficiently large $\mathrm{SNR} \gtrsim 20$, which is likely to have a small effect on the results presented here.

Finally, mass measurements critically rely on measuring eccentricity and correctly identifying a forest of lines with characteristic frequency separations (if one considers the properties of the signals in Fourier space) as coming from a single source in an eccentric orbit, rather than several distinct circular binaries all sharing the same sky locations. 
This aspect will need careful attention in the implementation of the \emph{global fit} which is one of the most challenging aspects of the LISA mission.

\section*{Acknowledgments}

We thank Diganta Bandopadhyay and Christopher~J.~Moore for helpful comments on the manuscript. 
A.K., H.M., and A.V. acknowledge the support of the UK Space Agency, Grant No. ST/V002813/1 and UKRI971. 
P.K. acknowledges support from STFC grant ST/V005677/1 and ST/W000946/1. 
A.V. acknowledges the support of the Royal Society and Wolfson Foundation. 
A.V. thanks the Flatiron Institute for hospitality while a portion of this research was carried out. 
The Flatiron Institute is a division of the Simons Foundation. 
Computational resources used for this work were provided by the University of Birmingham’s BlueBEAR High Performance Computing facility.

\textit{Software:} This work has made use of 
\texttt{numpy}~\citep{numpy:2020},
\texttt{scipy}~\citep{scipy:2020},
\texttt{astropy}~\citep{astropy:2018},
\texttt{nessai}~\citep{nessai},
\texttt{matplotlib}~\citep{matplotlib:2007}, 
and 
\texttt{seaborn}~\citep{seaborn:2021}.

\bibliography{refs}{}
\bibliographystyle{aasjournal}

\appendix

\section{Additional tables and plots}
\label{app:tables_box}

We provide illustrative results summarizing selected injected and recovered parameters in Table~\ref{tab:results_full}. 
For these systems one can measure (any choice of) two independent mass parameters, and we provide results for $M$, ${\cal M}$, $m_1$, and $m_2$. 
In addition, one can also measure the distance to the source, $D$.
Three examples of each source class are shown in Table~\ref{tab:results_full}.
As described in the main text in Table~\ref{tab:measure}, the total number of systems for which one can measure two independent mass parameters are $13$, $10$, and $49$ for BHBH, BHNS, and NSNS systems, respectively. 
The full tables of injected and recovered parameters for each system with measured mass parameters are included in machine-readable format in the accompanying data release at \url{doi.org/10.5281/zenodo.15704561}.

\begin{figure*}[h!]
    \centering 
    \includegraphics[width=0.245\linewidth]{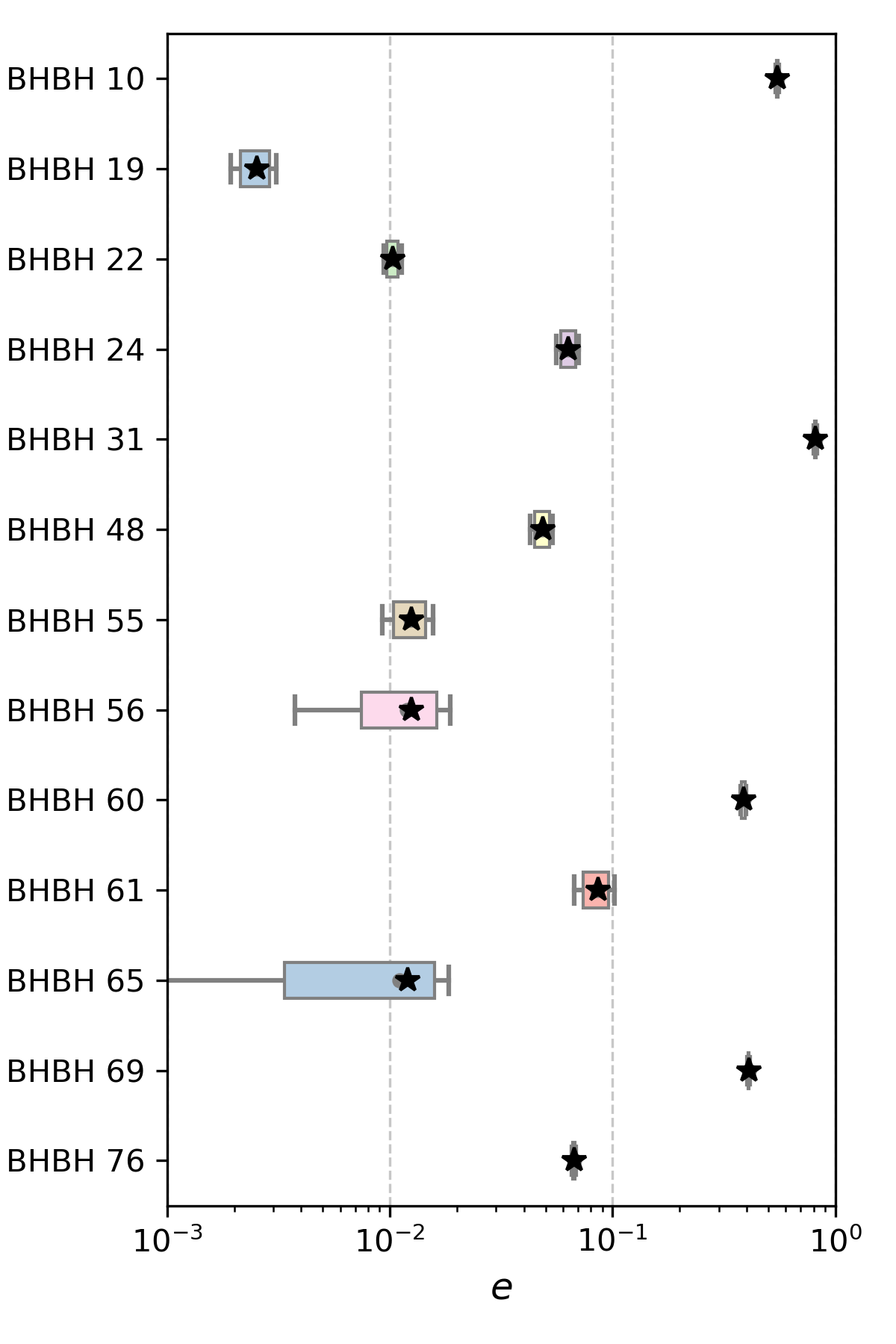}
    \includegraphics[width=0.245\linewidth]{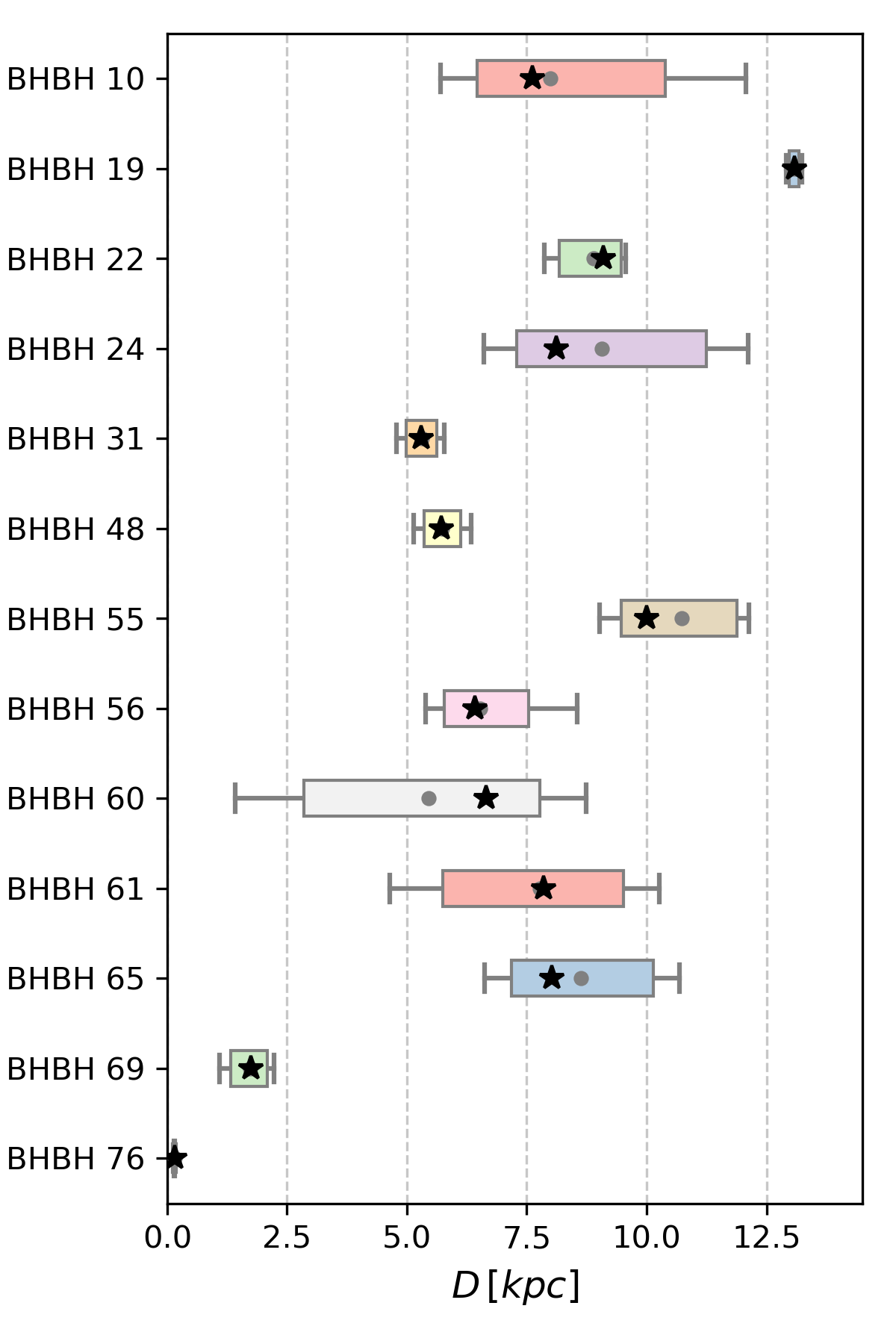}
    \includegraphics[width=0.245\linewidth]{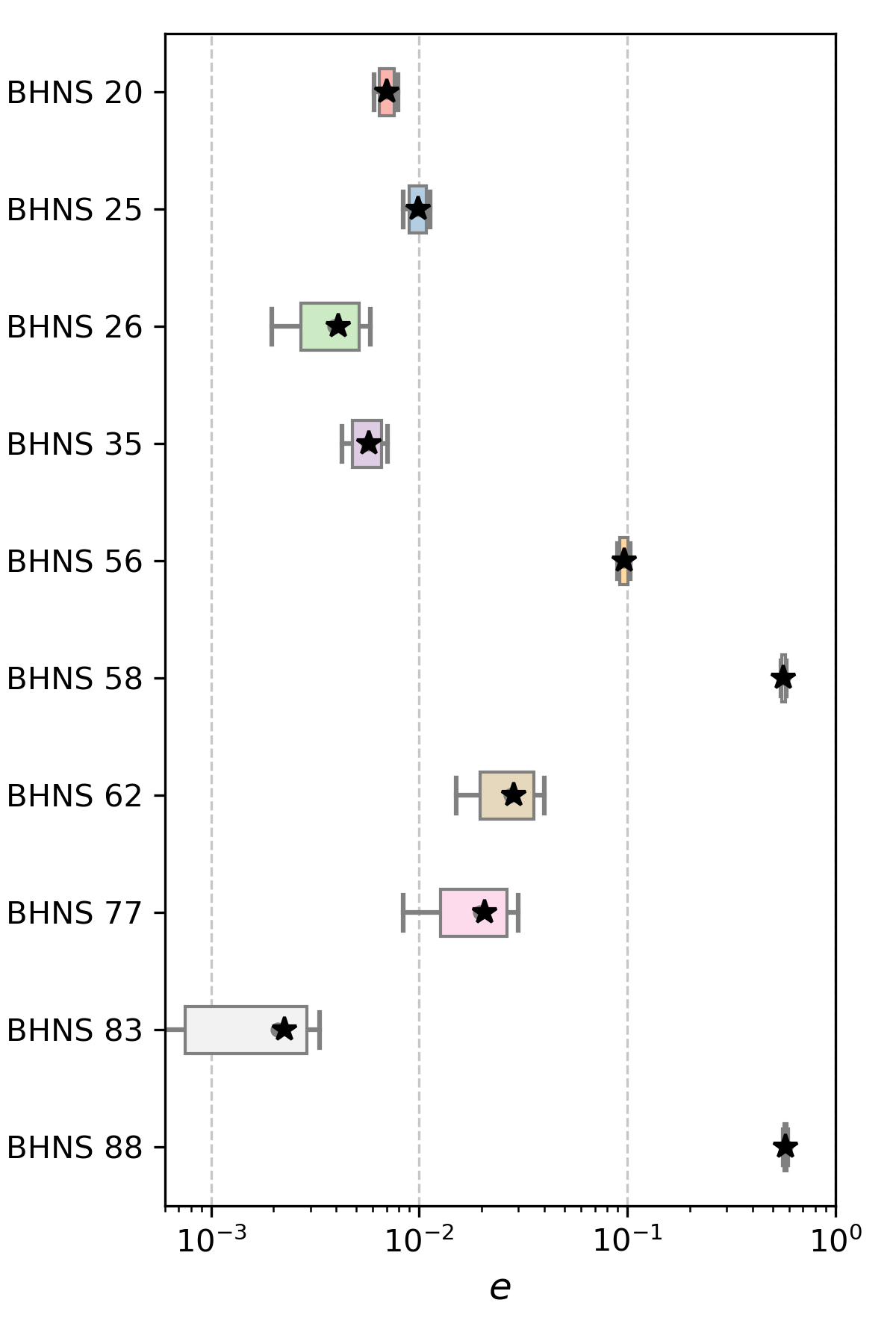}
    \includegraphics[width=0.245\linewidth]{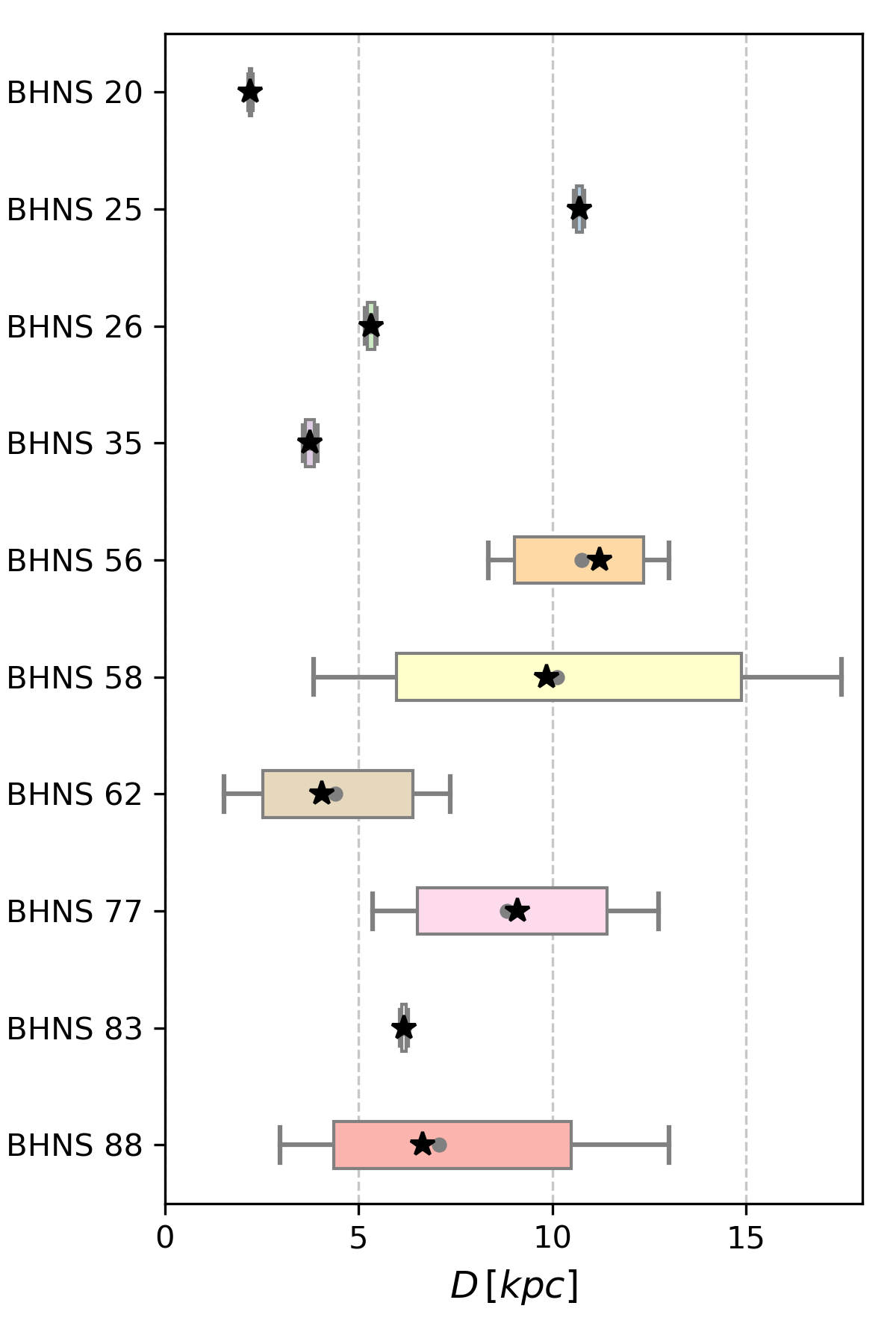}
    \vspace{-10pt} 
    \caption{
    Box plots showing the marginalised posterior distribution functions of the eccentricity $e$ and distance $D$ of the 13 BHBHs (left) and 10 BHNS (right) for which mass parameters can be measured (see also Table~\ref{tab:results_full} and the data release).
    On the vertical axis we show the injection ID number. 
    Coloured boxes show the symmetric $90\%$ probability interval and thin grey lines show the symmetric $99\%$ probability interval. 
    Dots denote the median value of the PDF. Stars correspond to the value of the injection.}
    \label{fig:de_box_BH}
\end{figure*}

\begin{figure*}[h!]
    \centering
    \includegraphics[width=0.25\linewidth]{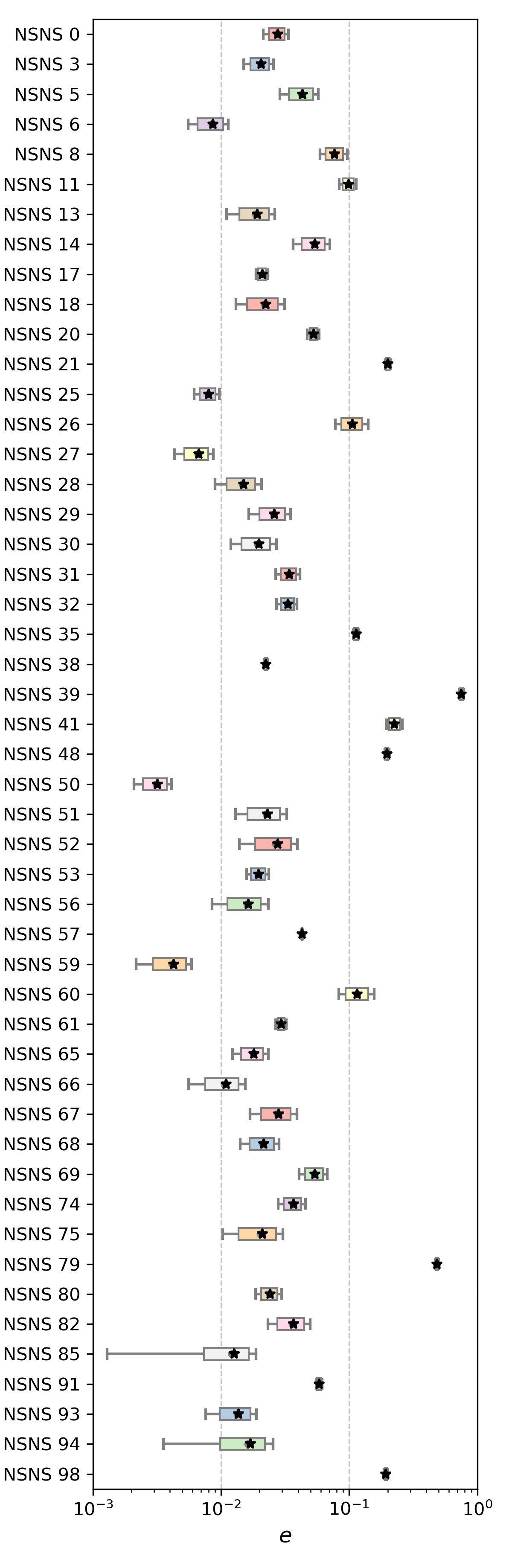}
    \includegraphics[width=0.25\linewidth]{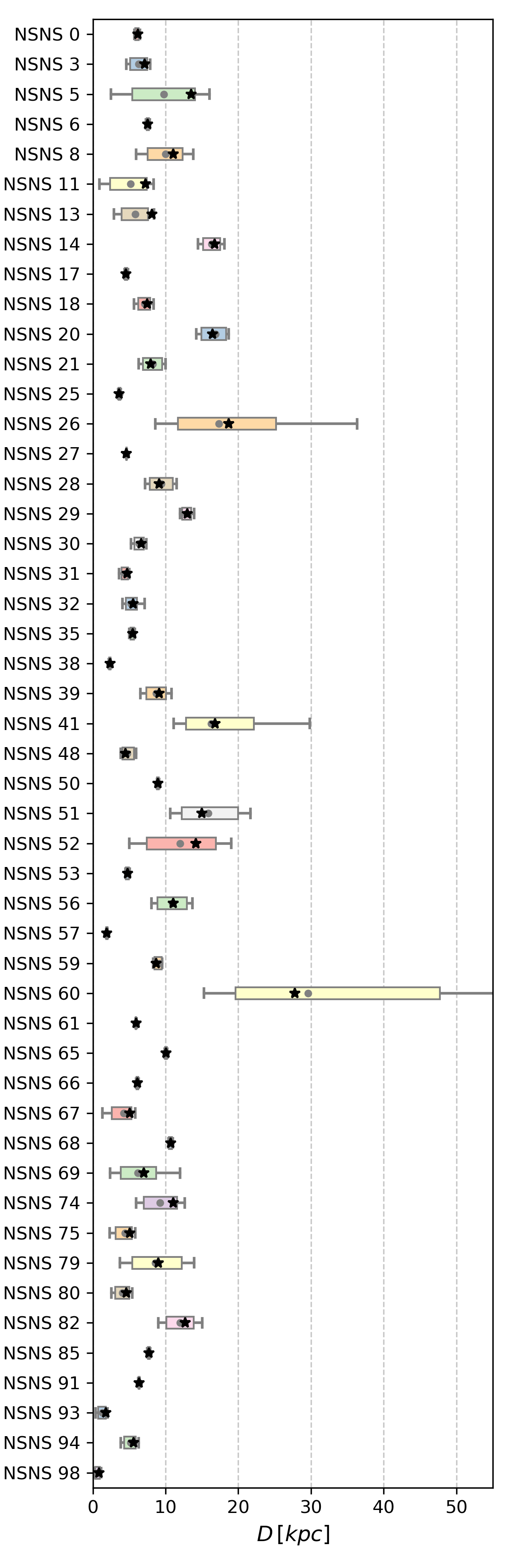}
    \vspace{-10pt} 
    \caption{Same as Fig.~\ref{fig:de_box_BH} for the 49 NSNSs for which mass parameters can be measured (see also Table~\ref{tab:results_full} and the data release).}
    \label{fig:de_box_BNS}
\end{figure*}

We provide the associated box plots of the marginalised 1D posterior distributions for the distance, $D$, and eccentricity, $e$, in Fig.~\ref{fig:de_box_BH} for BHBHs and BHNSs, and Fig.~\ref{fig:de_box_BNS} for NSNSs. 
They complement the box plots for the mass parameters shown in Figs.~\ref{fig:m_box_BHs} and~\ref{fig:m_box_BNSs}, respectively.

\begin{figure*}[htb!]
    \centering
    \includegraphics[width=0.95\textwidth]{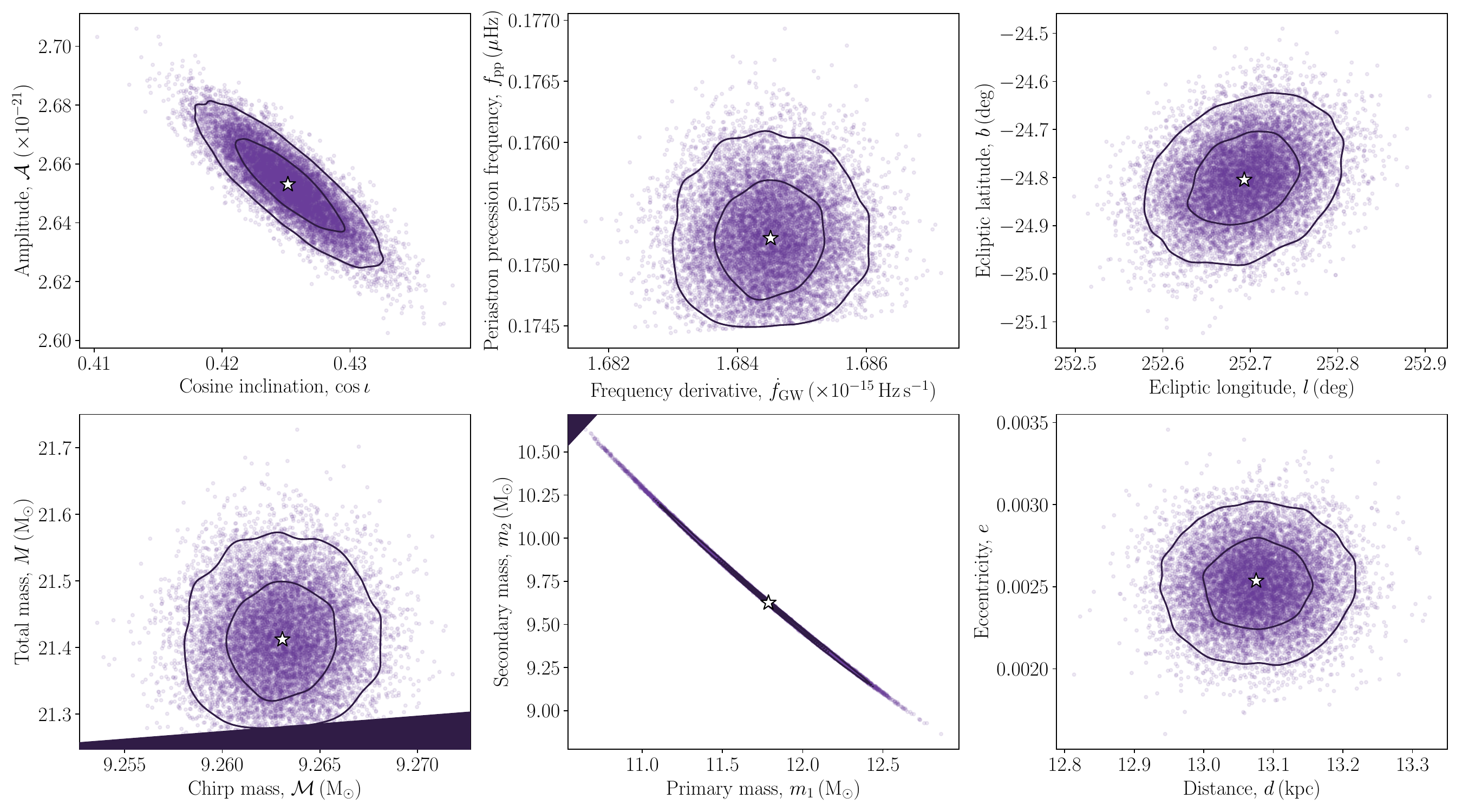}
    \caption{
    An example posterior distribution for BHBH ID-$19$.
    This signal has ${\rm SNR}=386$ and $f_{\rm GW}=1.7\,{\rm m Hz}$
    Each panel shows a two-dimensional posterior distribution. 
    Top row (from left to right): amplitude and cosine of inclination; periastron precession frequency and frequency derivative; and ecliptic latitude and ecliptic longitude. 
    Bottom row (from left to right): total mass and chirp mass; component masses; and eccentricity and distance. 
    In each panel, the markers shows the posterior samples and the contours indicate the $50\%$ and $90\%$ credible areas. 
    The star indicates the position of the injection. 
    The shaded regions indicate unphysical regions of the parameter space.
    In the $m_1$--$m_2$ plot the region where $m_2>m_1$ is unphysical and is shaded out. 
    In the $\mathcal{M}$--$M$ plot, the corresponding region where $\eta>0.25$ is shaded. 
    }
    \label{fig:BHBH19}
\end{figure*}

\begin{figure*}[htb!]
    \centering
    \includegraphics[width=0.95\textwidth]{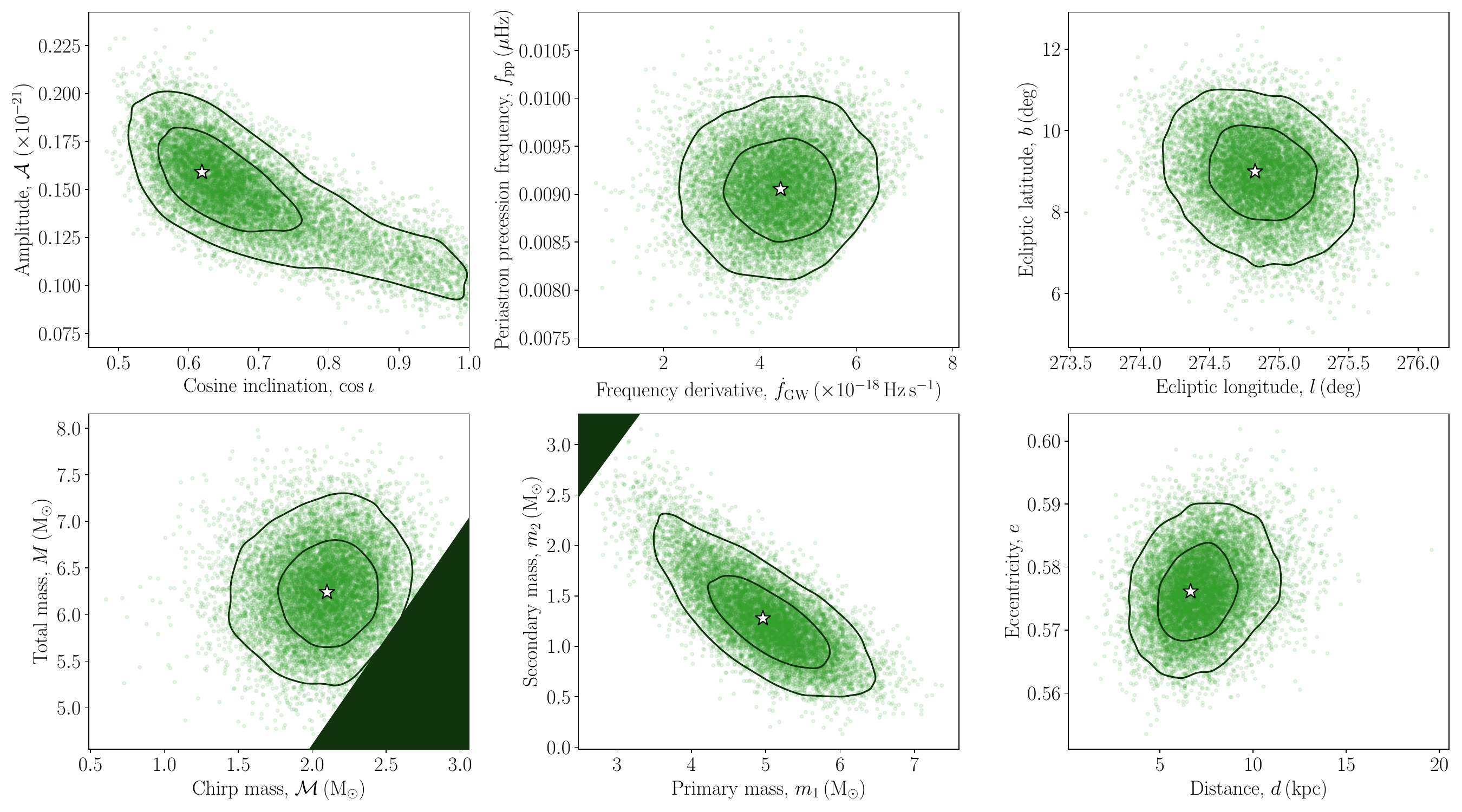}
    \caption{
    An example posterior for BHNS-88. 
    This signal has ${\rm SNR} = 48$ and $f_{\rm GW}=0.37\,{\rm mHz}$. 
    The panels are identical to Fig.~\ref{fig:BHBH19}.
    }
    \label{fig:BHNS88}
\end{figure*}

\begin{figure*}[htb!]
\centering
    \includegraphics[width=0.95\textwidth]{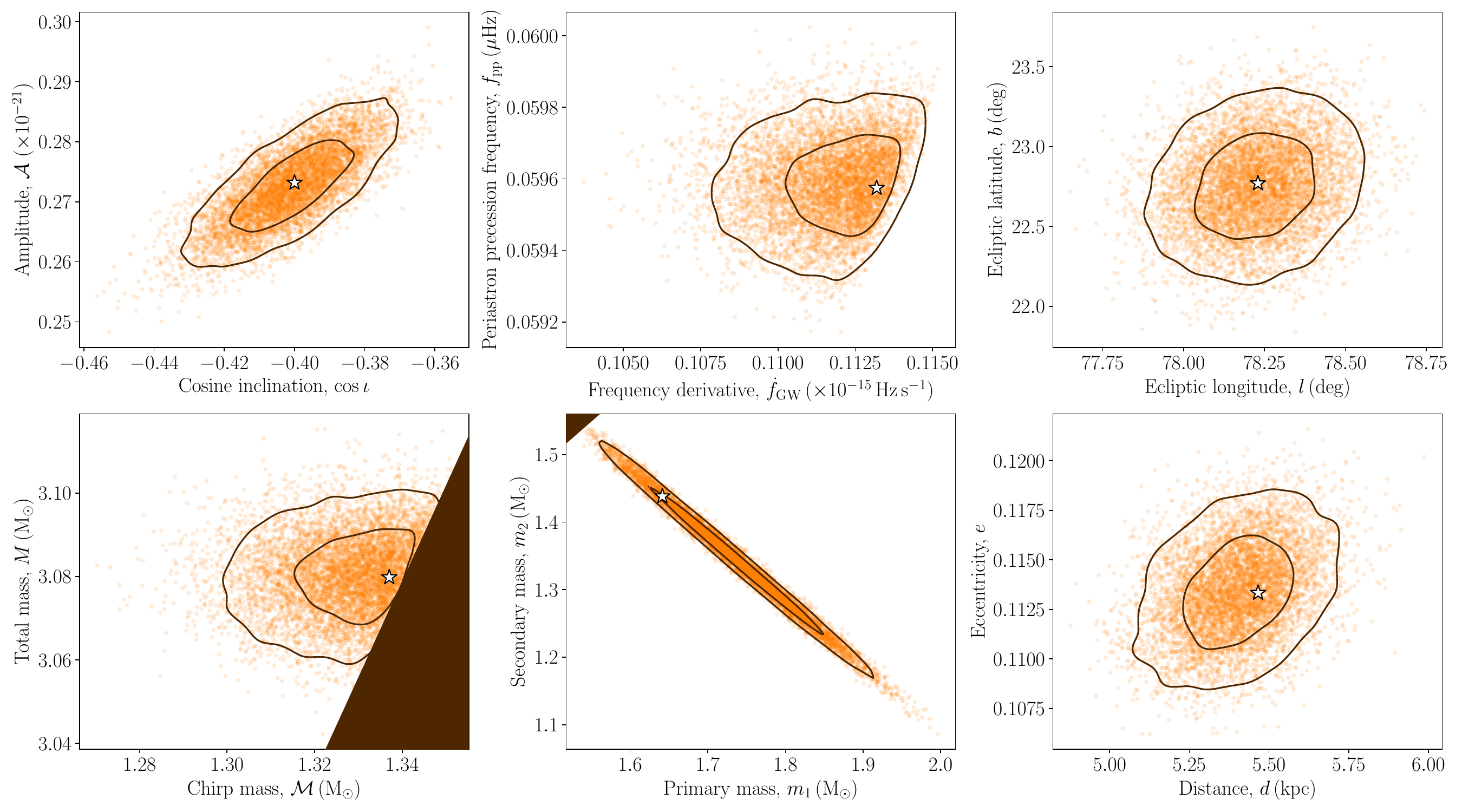}
    \caption{
    An example posterior for NSNS-$35$. 
    This signal has ${\rm SNR} = 88$ and $f_{\rm GW}=1.9\,{\rm mHz}$. 
    The panels are identical to Fig.~\ref{fig:BHBH19}.
    }
    \label{fig:NSNS35}
\end{figure*}

We also show two-dimensional PDFs for selected parameters---GW amplitude and inclination angle of the binary, $({\cal A}, \cos\iota)$, GW frequency derivative and precession frequency $(\dot{f}_\mathrm{GW}, f_\mathrm{PP})$, ecliptic longitude and latitude, $(l, b)$; and on derived parameters (with the exception of the eccentricity $e$), chirp and total mass $({\cal M}, M)$, individual masses $(m_1, m_2)$, and distance and eccentricity, $(D, e)$---for one injection for each class of binaries.
Fig.~\ref{fig:BHBH19} refers to BHBH ID-19, which is a low-eccentricity ($e = 2.5\times 10^{-3}$) BHBH in the middle of the LISA frequency band ($f_\mathrm{GW} = 1.7\,\mathrm{mHz}$) with $\mathrm{SNR} = 386$.
Fig.~\ref{fig:BHNS88} refers to BHNS ID-88, which is a low-frequency BHNS with frequency $f_\mathrm{GW} = 0.37 \,\mathrm{mHz}$ and large eccentricity ($e = 0.58$) with $\mathrm{SNR} = 48$.
Finally, Fig.~\ref{fig:NSNS35} refers to NSNS ID-35, which is a NSNS with frequency $f_\mathrm{GW} = 1.9 \,\mathrm{mHz}$ and moderate eccentricity ($e = 0.11$) with $\mathrm{SNR} = 88$.
The injected and recovered parameters for each of the systems highlighted above are shown in Table~\ref{tab:results_full}.

\begin{longrotatetable}
{\begin{deluxetable*}{l cc cccc cc ccc cccc}
\tablecaption{
Injection and measured parameter values for a selection of DCO systems for which two independent mass parameters are measured.
Here we include three examples of each source type. 
From the $100$ injections of each source type, we find that two masses can be measured for $13$ binary black hole systems (BHBHs), $10$ black hole--neutron star systems (BHNS), and $49$ binary neutron star systems (NSNS)
This table is available in its entirety in machine-readable form in the online article at \url{https://doi.org/10.3847/1538-4357/ae9c34}.
\label{tab:results_full}}
\tablewidth{0pt}
\tabletypesize{\small}
\tablehead{
  \colhead{} &
  \multicolumn{8}{c}{Injection values} &
  \multicolumn{7}{c}{Measured values} \\
  \cline{2-9} \cline{10-16} \\[-6pt]
  \colhead{ID} &
  \colhead{$f_{\mathrm{GW}}$} &
  \colhead{SNR} &
  \colhead{$e$} &
  \colhead{$D$} &
  \colhead{$m_1$} &
  \colhead{$m_2$} &
  \colhead{$M$} &
  \colhead{$\mathcal{M}$} &
  \colhead{$e$} &
  \colhead{$D$} &
  \colhead{$\Delta\Omega$} &
  \colhead{$m_1$} &
  \colhead{$m_2$} &
  \colhead{$M$} &
  \colhead{$\mathcal{M}$} \\
  \colhead{} &
  \colhead{(mHz)} &
  \colhead{} &
  \colhead{} &
  \colhead{(kpc)} &
  \colhead{($M_\odot$)} &
  \colhead{($M_\odot$)} &
  \colhead{($M_\odot$)} &
  \colhead{($M_\odot$)} &
  \colhead{} &
  \colhead{(kpc)} &
  \colhead{(deg$^2$)} &
  \colhead{($M_\odot$)} &
  \colhead{($M_\odot$)} &
  \colhead{($M_\odot$)} &
  \colhead{($M_\odot$)}
}
\startdata
\multicolumn{3}{l}{BHBH binaries}\\
BHBH\,19 & $1.7$  & $386$ & $0.003$ & $13.1$ & $11.8$ & $9.6$ & $21.0$ & $9.3$ &
  $0.0025_{-0.0004}^{+0.0004}$ & $13.08_{-0.10}^{+0.11}$  & $0.055$  &
  $11.80_{-0.62}^{+0.49}$  & $9.62_{-0.36}^{+0.52}$  & $21.41_{-0.10}^{+0.12}$   & $9.263_{-0.004}^{+0.004}$ \\
BHBH\,31 & $0.17$ & $209$ & $0.82$  & $5.3$  & $7.3$  & $3.0$ & $10.3$ & $4.0$ &
  $0.815_{-0.008}^{+0.008}$  & $5.31_{-0.32}^{+0.31}$   & $0.460$  &
  $7.30_{-0.65}^{+0.62}$   & $3.05_{-0.46}^{+0.58}$  & $10.37_{-0.68}^{+0.69}$  & $4.04_{-0.35}^{+0.36}$   \\
BHBH\,69 & $0.42$ & $71$  & $0.41$  & $1.7$ & $8.0$  & $3.3$ & $11.3$ & $4.4$ &
  $0.408_{-0.003}^{+0.003}$ & $1.74_{-0.40}^{+0.35}$ & $1.045$  &
  $8.0_{-1.4}^{+1.0}$      & $3.2_{-1.0}^{+1.4}$     & $11.26_{-0.17}^{+0.16}$  & $4.34_{-0.63}^{+0.46}$   \\
\multicolumn{1}{c}{\vdots} \\
\multicolumn{3}{l}{BHNS binaries}\\
BHNS\,26 & $2.1$  & $191$ & $0.004$ & $5.3$  & $4.6$  & $1.3$ & $5.8$  & $2.0$ &
  $0.004_{-0.001}^{+0.001}$  & $5.32_{-0.09}^{+0.10}$   & $0.193$  &
  $4.55_{-0.15}^{+0.15}$   & $1.28_{-0.04}^{+0.04}$  & $5.83_{-0.11}^{+0.11}$   & $2.02_{-0.01}^{+0.01}$   \\
BHNS\,62 & $0.90$ & $41$  & $0.03$  & $4.1$  & $6.5$  & $2.2$ & $8.7$  & $3.2$ &
  $0.028_{-0.008}^{+0.008}$  & $4.4_{-1.9}^{+2.0}$      & $10.723$ &
  $6.6_{-1.5}^{+1.2}$      & $2.1_{-1.0}^{+1.3}$     & $8.73_{-0.51}^{+0.52}$   & $3.15_{-0.85}^{+0.55}$   \\
BHNS\,88 & $0.37$ & $48$  & $0.58$  & $6.6$  & $5.0$  & $1.3$ & $6.2$  & $2.1$ &
  $0.576_{-0.010}^{+0.010}$  & $7.1_{-2.7}^{+3.4}$      & $4.543$  &
  $5.0_{-1.2}^{+1.0}$      & $1.26_{-0.56}^{+0.78}$  & $6.26_{-0.74}^{+0.80}$   & $2.09_{-0.51}^{+0.41}$   \\
\multicolumn{1}{c}{\vdots} \\
\multicolumn{3}{l}{NSNS binaries}\\
NSNS\,35 & $1.9$  & $88$  & $0.11$  & $5.5$  & $1.6$  & $1.4$ & $3.1$  & $1.3$ &
  $0.113_{-0.004}^{+0.004}$  & $5.40_{-0.24}^{+0.24}$   & $0.569$  &
  $1.74_{-0.13}^{+0.14}$   & $1.34_{-0.13}^{+0.13}$  & $3.08_{-0.02}^{+0.02}$   & $1.33_{-0.02}^{+0.01}$   \\
NSNS\,59 & $4.6$  & $554$ & $0.004$ & $8.7$  & $1.3$  & $1.3$ & $2.6$  & $1.1$ &
  $0.004_{-0.001}^{+0.001}$  & $9.04_{-0.55}^{+0.44}$   & $0.009$  &
  $1.38_{-0.07}^{+0.08}$   & $1.18_{-0.06}^{+0.06}$  & $2.566_{-0.008}^{+0.020}$ & $1.1128_{-0.0003}^{+0.0003}$ \\
NSNS\,79 & $0.72$ & $26$  & $0.49$  & $9.0$  & $1.6$  & $1.3$ & $2.9$  & $1.3$ &
  $0.485_{-0.010}^{+0.009}$  & $8.6_{-3.2}^{+3.6}$      & $2.142$  &
  $2.06_{-0.43}^{+0.43}$   & $0.87_{-0.40}^{+0.41}$  & $2.93_{-0.16}^{+0.17}$   & $1.14_{-0.26}^{+0.14}$   \\
\multicolumn{1}{c}{\vdots} \\
\enddata
\tablecomments{Injection values of eccentricity $e$, luminosity distance $D$, and component,
total, and chirp masses ($m_1$, $m_2$, $M$, $\mathcal{M}$) are listed alongside the corresponding
median recovered values with $90\%$ credible intervals. $\Delta\Omega$ is the $90\%$
credible sky-localisation area.}
\end{deluxetable*}}
\end{longrotatetable}

\end{document}